\documentclass[12pt]{article}
\pdfoutput=1
\usepackage{epsf,amsfonts,hyperref}
\usepackage[dvips]{color}
\usepackage{amsmath}
\usepackage{graphicx}
\renewcommand{\appendix}[1]{
    \setcounter{equation}{0}
    \renewcommand{\thesection}{\Alph{section}}
    \section{Appendix: \protect\indent #1}
}

\newcommand\encadremath[1]{\vbox{\hrule\hbox{\vrule\kern8pt
\vbox{\kern8pt \hbox{$\displaystyle #1$}\kern8pt}
\kern8pt\vrule}\hrule}}
\def\enca#1{\vbox{\hrule\hbox{
\vrule\kern8pt\vbox{\kern8pt \hbox{$\displaystyle #1$}
\kern8pt} \kern8pt\vrule}\hrule}}

\newcommand\figureframex[3]{
\begin{figure}[bth]
\hrule\hbox{\vrule\kern8pt
\vbox{\kern8pt \vbox{
\begin{center}
{\mbox{\epsfxsize=#1.truecm\epsfbox{#2}}}
\end{center}
\caption{#3}
}\kern8pt}
\kern8pt\vrule}\hrule
\end{figure}
}
\newcommand\figureframey[3]{
\begin{figure}[bth]
\hrule\hbox{\vrule\kern8pt
\vbox{\kern8pt \vbox{
\begin{center}
{\mbox{\epsfysize=#1.truecm\epsfbox{#2}}}
\end{center}
\caption{#3}
}\kern8pt}
\kern8pt\vrule}\hrule
\end{figure}
}

\makeatletter
\@addtoreset{equation}{section}
\makeatother
\newtheorem{theorem}{Theorem}[section]

\newtheorem{remark}{Remark}[section]
\newtheorem{proposition}{Proposition}[section]
\newtheorem{lemma}{Lemma}[section]
\newtheorem{corollary}{Corollary}[section]
\newtheorem{definition}{Definition}[section]
\def\br{\begin{remark}\rm\small}
\def\er{\end{remark}}
\def\bt{\begin{theorem}}
\def\et{\end{theorem}}
\def\bd{\begin{definition}}
\def\ed{\end{definition}}
\def\bp{\begin{proposition}}
\def\ep{\end{proposition}}
\def\bl{\begin{lemma}}
\def\el{\end{lemma}}
\def\bc{\begin{corollary}}
\def\ec{\end{corollary}}
\def\beaq{\begin{eqnarray}}
\def\eeaq{\end{eqnarray}}

\newcommand{\beq}{\begin{equation}}
\newcommand{\eeq}{\end{equation}}
\newcommand{\bea}{\begin{eqnarray}}
\newcommand{\eea}{\end{eqnarray}}

 \newcommand{\Tr}{{\,\rm Tr}\:}

\newcommand{\td}[1]{{\tilde{#1}}}

\newcommand{\om}{\omega}

\newcommand{\Pint}{{\int\kern -1.em -\kern-.25em}}

\renewcommand{\Re}{{\mathrm{Re}}}
\renewcommand{\Im}{{\mathrm{Im}}}

\newcommand\Res{\mathop{{\rm Res}}}

\begin{document}
%=============================Page de titre==============%\date{??}

\sloppy

%\maketitle

\pagestyle{empty}
%\hfill SPhT-T05/045
\addtolength{\baselineskip}{0.20\baselineskip}
\begin{center}
\vspace{26pt}
{\large \bf {Double scaling limits of random matrices and minimal $(2m,1)$ models: the merging of two cuts in a degenerate case}}
\vspace{20pt}
\newline
{\sl O.\ Marchal}\hspace*{0.05cm}\footnote{ E-mail: olivier.marchal@polytechnique.org }\\
D\'{e}partement de math\'{e}matiques et de statistique\\
Universit\'{e} de Montr\'{e}al, Canada\\
Institut de physique th\'{e}orique\\
F-91191 Gif-sur-Yvette Cedex, France.\\
\vspace{6pt}
{\sl M.\ Cafasso}\hspace*{0.05cm}\footnote{ E-mail: cafasso@crm.umontreal.ca }\\
Centre de recherches math\'{e}matiques\\
Concordia University\\
Montr\'{e}al, Canada\\

\vspace{6pt}
\end{center}

\vspace{20pt}
\begin{center}
{\bf Abstract}: 
In this article, we show that the double scaling limit correlation functions of a random matrix model when two cuts merge with degeneracy $2m$ (i.e. when $y\sim x^{2m}$ for arbitrary values of the integer $m$) are the same as the determinantal formulae defined by conformal $(2m,1)$ models. Our approach follows the one developed by Berg\`{e}re and Eynard in \cite{BergereEynard} and uses a Lax pair representation of the conformal $(2m,1)$ models (giving Painlev\'e II integrable hierarchy) as suggested by Bleher and Eynard in \cite{BleherEynard}. In particular we define Baker-Akhiezer functions associated to the Lax pair to construct a kernel which is then used to compute determinantal formulae giving the correlation functions of the double scaling limit of a matrix model near the merging of two cuts. 
\end{center}

\tableofcontents

%-----------------------------ABSTRACT--------------------------------------
\vspace{26pt}
\pagestyle{plain}
\setcounter{page}{1}

%*********************************************************************
%==================== ARTICLE =======================================%******************************************

\section{Introduction}

It has been known for a long time that the study of random matrix models in different scaling limits gives rise to a great number of well-known integrable equations; both PDEs of solitonic type (KdV and, more generally, Gelfand-Dikii equations) and ODEs arising from isomonodromic systems (like Painlev\'e equations). A key idea in these studies is the notion of spectral curves attached to algebraic equations $P(x,y)=0$. The genus of the curve gives the number of intervals on which the eigenvalues of the matrices will accumulate when their size tends to infinity. It is well known that, in the generic case, the curve behaves like $y\sim \sqrt{x-a}$ near a branchpoint $a$ (an extremity of an interval); the appropriate double scaling limit gives the celebrated Airy kernel in connection with the $(1,2)$ minimal model. But it may happen by taking a fine-tuned limit (see for instance \cite{BergereEynard, BleherIts}), that the behavior near a branchpoint differs from the generic case and takes the form of $y^q \sim (x-a)^p$. In such a case, it is expected that the double scaling limit is related to the conformal $(p,q)$ minimal model. In their articles \cite{BergereEynard} and \cite{determinantalformulae}, the authors opened the way to rigorous mathematical proofs in order to establish that the correlation functions of the double-scaling limit of a matrix model are the same as the ones defined by determinantal formulae arising from $(p,q)$ models. In their articles, they apply this method to all $(2m+1,2)$ models, i.e. suitable limits of matrix models where the spectral curve behaves like $y^2 \sim x^{2m+1}$ near an endpoint. In this article, we will use the same method for the $(2m,1)$ case which corresponds to a point where two cuts are merging with a degeneracy $2m$. For a generic merging, i.e. $m=1$ it has been proven in \cite{BleherEynard} that the suitable double scaling limit of the matrix model is connected to the Painlev\'e II equation. Some similar results have been established with the study of a suitable Riemann-Hilbert problem. For example the case of an even-quartic polynomial has been studied in \cite{BleherIts}. It would be also interesting to derive similar results, for these kernels, as the ones proved in \cite{CIK}. Here, using the approach of \cite{BergereEynard}, we find, as expected, that the correlation functions of the double scaling limit of the merging of two cuts with degeneracy $2m$ are expressed through the Lax system of the Painlev\'{e} II hierarchy (see \cite{CJM} and \cite{MazzoccoMo}).

\section{Double scaling limit in random matrices: the merging of two cuts}
\subsection{Hermitian matrix models and equilibrium density} 
It is well known in the literature that the study of the Hermitian matrix model with partition function:
\beq \label{HermIntegral} Z_N=\int_{\mathbb{H}_N} \exp(-N\Tr(V(M)))dM\eeq
with an even polynomial potential
\beq \label{defPotential}
V(x)=\sum_{i=1}^{2d} t_i x^i
\eeq
can be reduced to an eigenvalue problem: $\lambda=\{ \lambda_j,j=1,...,N\}$ for the matrix $M$ with distribution:
\beq \label{DiagProblem}
\td{Z}_N=\int_{\mathbb{R}^N} \exp\left(2\sum_{1\leq j<k\leq N} \log|\lambda_j-\lambda_k|-N\sum_{i=1}^N V(\lambda_j)\right)d\lambda_1\ldots d\lambda_N
\eeq
When $N \to \infty$, the distribution of the eigenvalues on the line $d\nu_N(x)=\rho_N(x)dx$ is defined (in the distribution theory sense) by the formula
\beq \label{distribdensity}
\int_{\mathbb{R}}\phi(x)d\nu_N(x)=\frac{1}{\td{Z}_N}\int_{\mathbb{R}^N}\left(\frac{1}{N}\sum_{j=1}^n \phi(\lambda_j)\right) \exp\left(2\sum_{1\leq j<k\leq N} \log|\lambda_j-\lambda_k|-N\sum_{i=1}^N V(\lambda_j)\right).
\eeq
For any test function $\phi(x)$ there is a weak limit $d\nu_\infty (x):=\underset{N \to \infty}{\lim} d\nu_N(x)$ which is the same as the equilibrium density $d\nu_{eq}(x)$ given by the limit of the empirical density:
\beq \label{empdistribution}
d \nu_{eq}(x)=\mathop{\lim}_{N \to \infty} \frac{1}{N} \sum_{j=1}^N \delta(x-\lambda_j) 
\eeq
For details about the existence of the distribution limits, the equality between the equilibrium density $d\nu_{eq}(x)$ and $d\nu_\infty(x)$ and the following characterizations we refer the reader to \cite{BPS}, \cite{JOH}. Nowadays, many properties of the equilibrium density are known. For example, we know \cite{MehtaBook} that the equilibrium density is supported by a finite number of intervals $[a_j,b_j], \, j=1,\dots,q$ and that it is absolutely continuous with respect to the Lebesgue measure:
\beq d\nu_{eq}(x)=\rho(x)dx=\frac{1}{2i\pi} h(x) R^{1/2}(x)dx, \,\, \, R(x)dx=\prod_{j=1}^q \left((x-a_j)(x-b_j)\right)dx\eeq
where $h(x)$ is a polynomial of degree $2d-q-1$ and $R^{1/2}(x)$ is to be understood as the value on the upper cut of the principal sheet of the complex-valued function $R^{1/2}(z)$ with cuts on $J=\bigcup_{j=1}^q[a_j,b_j]$. In fact, the equilibrium density $d\nu_{eq}(x)$ is completely defined by the knowledge of the extremities $a_j$'s and $b_j$'s and the unknown coefficients of the polynomial $h(x)$. It has been proved \cite{Deift:1998p16} that such quantities are uniquely determined by the following set of equations:
\begin{enumerate}
 \item Connexion between $h(z)$ and the potential $V(z)$\footnote{Here and below we denote with Pol$(A(z))$ the polynomial part of the asymptotic expansion of $A(z)$ at infinity}:
\beq V'(z)=\underset{z \to \infty}{\text{Pol}}\left( h(z) R^{1/2}(z)\right) \label{Vconnexionh}\eeq
 \item Residue constraint:
\beq \label{ResidueConstraint}\Res_{z \to \infty} \left(h(z)R^{1/2}(z)\right)=-2\eeq
 \item Integrals constraints:
\beq \label{IntegralsConstraints}\int_{b_j}^{a_{j+1}}h(z)R^{1/2}(z)dz=0\, ,\,\, \forall j\in \{1,\dots,q-1\}
\eeq
\end{enumerate}
Note also that the relation between $h(z)$ and $V(z)$ (\ref{Vconnexionh}) can be inverted by:
\beq \label{hconnexionV} h(z)=\underset{z \to \infty}{\text{Pol}}\left(\frac{V'(z)}{R^{1/2}(z)}\right)\eeq

In theory, the previous set of equation is sufficient to determine the whole solution $d\nu_{eq}(x)$ but, practically, since the equations are highly non-linear, it becomes very hard to compute the unknown coefficients for two or more intervals or for potentials of degree higher than $4$. Moreover, in some exceptional situations, the previous set of equations has multiple solutions. In such situations, the good solution is determined by a positivity condition:
\beq \label{positivity} h(x)\geq0, \,\, \forall x \in J=\bigcup_{j=1}^q[a_j,b_j]\eeq
When $\forall x \in \bigcup_{j=1}^q]a_j,b_j[: \,\, h(x)>0$, the potential $V(x)$ and the equilibrium measure $d\nu_{eq}(x)$ are called \textit{regular}. Otherwise the equilibrium density is called \textit{singular} and the corresponding potential is called \textit{critical}, meaning that there is at least one point on $J$ where the equilibrium measure vanishes. For a regular potential, the situation can be summarized with the following picture:

\begin{center}
	\includegraphics[height=5cm]{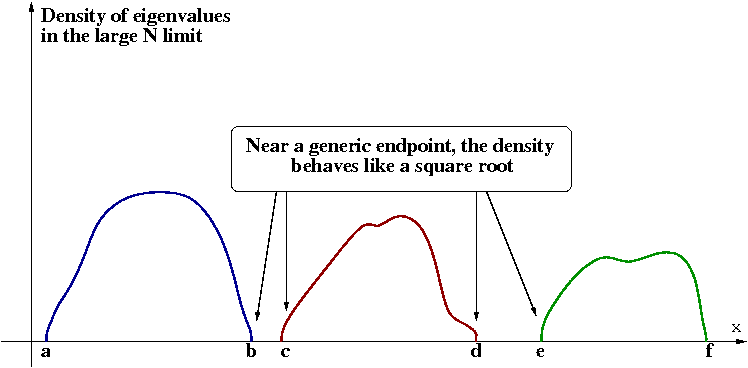}

\end{center}
	\underline{Figure 1:} Example of a typical eigenvalue density for a regular potential. The density is spread here in three intervals.
%\begin{figure}[h]
%\resizebox{0.8\textwidth}{!}{\input{DensitySeveralCuts.png}}
%\caption{Example of a typical eigenvalue density for a regular potential. The density is spread here in three intervals.}
%\end{figure}

\subsection{Singular densities for the $(2m,1)$ case}

In order to study what happens at a singular density, one embeds the potential $V(x)$ into a parametric family $V(x,t)$ so that for some $t=t_c$ the problem is at the critical potential: $V(x,t_c)=V_c(x)$. Then the interesting questions are to determine the asymptotics of the eigenvalues correlation functions when $t \to t_c$. Indeed for $t \neq t_c$ the potential is regular and we are in the situation depicted in the figure above. Therefore one can define $a_j(t)$, $b_j(t)$ and $h(x,t)$ determining completely the equilibrium density for $t\neq t_c$ and study their limits when $t \to t_c$.
In matrix models, it is often interesting to study a modified version of the integral (\ref{HermIntegral}) by introducing a parameter $T$ often referred as ``\textit{the temperature}'':
\beq \label{HermIntegralwithT} Z_N=\int_{\mathbb{H}_N} \exp(-\frac{N}{T}\Tr(V(M)))dM\eeq
It turns out that $T$ can be used as a parameter for the study of singular densities. In order to fit into our previous description, we need to introduce the following notation:
\beq V(x,T)=\frac{V(x)}{T}\eeq
To the study of the $(2m,1)$ model, we place ourselves in the case such that at $T=T_c$ the potential $V(x,T_c)=V_c(x)$ becomes singular (see formula (\ref{Potential}) below) and gives rise to a singular density defined by the following $2m$ singular density:
\beq \label{SingularDensity}
\rho(x,T_c)=\rho_c(x)=\frac{1}{2i\pi} (x-b\epsilon)^{2m}\sqrt{x^2-b^2}=\frac{1}{2i\pi}h_c(x)\sqrt{x^2-b^2}\eeq
with $\epsilon \in ]-1,1[$ representing the position of the singular point in the interval $]-b,b[$ supporting the distribution.
For $T\neq T_c$, we assume that the density is supported by two intervals $]a_1(T),b_1(T)[$ and $]a_2(T),b_2(T)[$ and define (note the normalization with $\frac{1}{T}$):
\beq \label{Density}
\rho(x,T)= \frac{1}{2i\pi T} h(x,T)\sqrt{(x-a_1(T))(x-b_1(T))(x-a_2(T))(x-b_2(T))}\eeq

Note that in order to recover our singular density at $T=T_c$ we must have:
\begin{center}
\begin{enumerate}
 \item $a_1(T)\underset{T\to T_c} {\to} -b$
 \item $b_1\underset{T\to T_c} {\to} b\epsilon$
 \item $a_2(T)\underset{T\to T_c} {\to} b\epsilon$
 \item $b_2(T) \underset{T\to T_c} {\to} b$
 \item $(x-b\epsilon)h(x,T)\underset{T\to T_c} {\to} T h_c(x)$
\end{enumerate}
\end{center}

The previous assumptions correspond to the merging to two cuts with degeneracy $2m$ (order of the singularity). The most general case would be a singular point $a$ with $\rho_c^q(x)\underset{T\to T_c}{\sim}(x-a)^p, \, (p,q) \in \mathbb{N}^2$, which is expected to correspond to the $(p,q)$ minimal model (for $q>2$ we are speaking about multi-matrix models). In our case the situation can be summarized with the following picture:\newpage

\begin{figure}[h]
\resizebox{0.8\textwidth}{!}{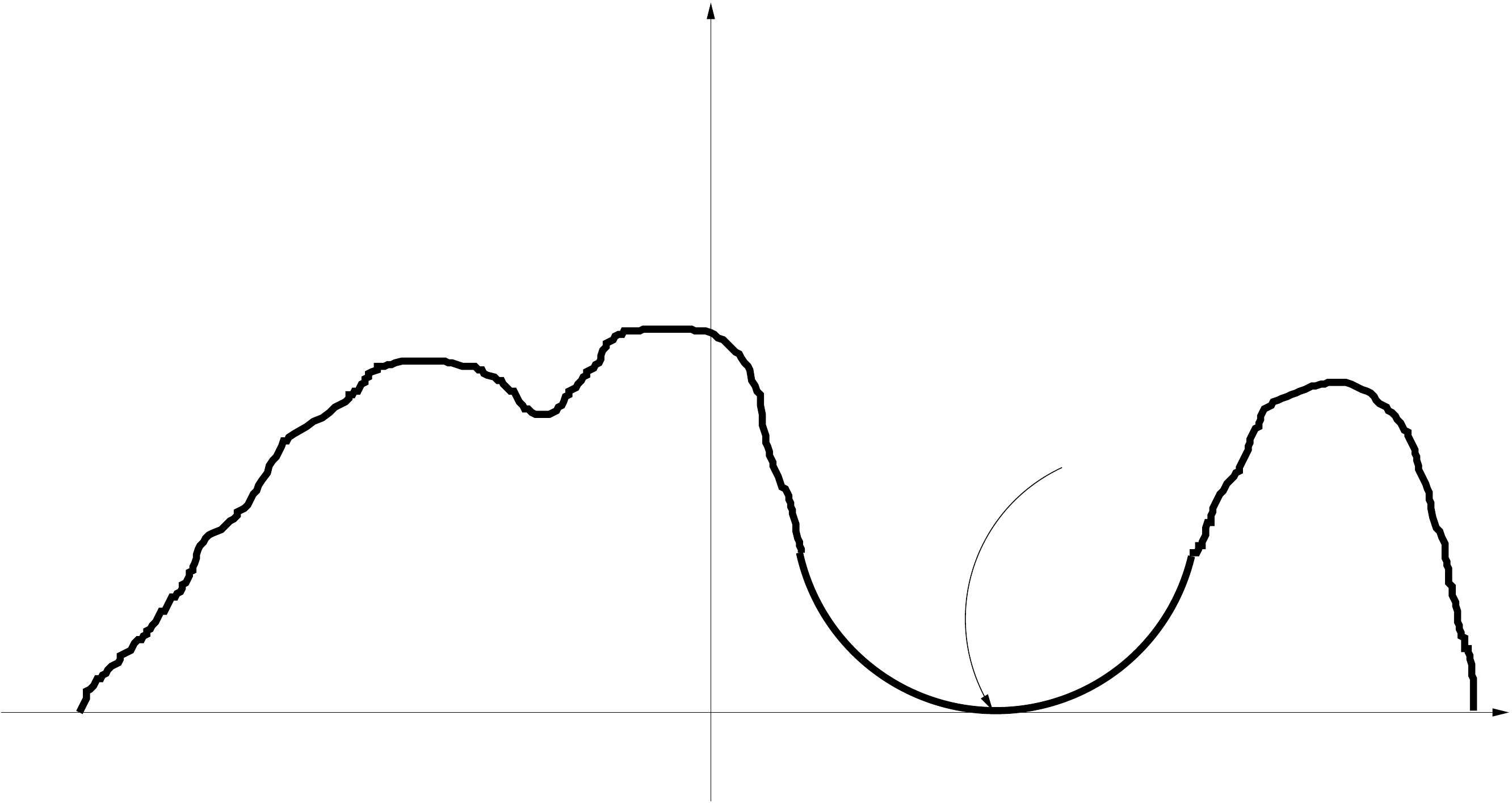}
%\label{FigDensity}
\end{figure}

\underline{Figure 2:} Example of a critical eigenvalue density; at the point $b\epsilon$ the density is singular and it behaves like $(x-b\epsilon)^{2m}$.\\

%\begin{center}
%	\includegraphics[height=6cm]{Degeneratemerging.png}
%
%	\underline{Figure 2}: Example of a critical eigenvalue density for a critical potential. At point $c$, the density is singular and behaves like $x^{2m}$

%\bigskip
%\bigskip

%\includegraphics[height=6cm]{DegeneratemergingSym.png}

%	\underline{Figure 3}: Example of a critical eigenvalue density for a critical even potential. At the origin, the density is singular and behaves like $x^{2m}$
%\end{center}

In \cite{BleherEynard}, the authors studied the case $m=1$ in details and conjectured some connections with Painlev\'{e} II hierarchy for higher $m$.

\subsection{Double scaling limits in matrix models}

In the study of matrix models, one is usually interested in the following functions called \textit{resolvents}:
\bea \label{Corrfunctions}w_n(x_1,\dots,x_n)&:=&\big{<} \Tr\left(\frac{1}{x_1-M}\right)\dots \Tr\left(\frac{1}{x_n-M}\right)\big{>}\cr
&=&\big{<}\sum_{i_1,\dots,i_n} \left(\frac{1}{x_1-\lambda_{i_1}}\right)\dots \left(\frac{1}{x_n-\lambda_{i_n}}\right)\big{>}
\eea
and in their cumulants, also known as \textit{correlation functions}:
\bea \label{Corrfunctions2}\hat{w}_n(x_1,\dots,x_n)&:=&\big{<} \Tr\left(\frac{1}{x_1-M}\right)\dots \Tr\left(\frac{1}{x_n-M}\right)\big{>}_{c}\cr
&=&\big{<}\sum_{i_1,\dots,i_n} \Tr\left(\frac{1}{x_1-\lambda_{i_1}}\right)\dots \Tr\left(\frac{1}{x_n-\lambda_{i_n}}\right)\big{>}_{c}
\eea
Here, the brackets stand for the integration relatively to the probability measure $Z_N^{-1}d\nu_N(x)$, the $\lambda_i$'s are the eigenvalues of the matrices and the index $c$ stands for the cumulants part defined as follows: 
\bea <A_1> &=& <A_1>_c \cr
<A_1 A_2> &=& <A_1 A_2>_c + <A_1>_c <A_2>_c \cr
<A_1 A_2 A_3> &=& <A_1 A_2 A_3>_c + <A_1 A_2>_c <A_3> + <A_1 A_3>_c <A_2> \cr
&&+ <A_2 A_3>_c <A_3> + <A_1> <A_2> <A_3>\cr
<A_1\dots A_n>&=&<A_J>=\sum_{k=1}^n\sum_{I_1 \bigsqcup I_2 \dots \bigsqcup I_k=J}\prod_{i=1}^k <A_{I_i}>_c.\cr\nonumber
\eea 
The joint density correlation functions $\rho_n(x_1,\dots,x_n)$ can easily be deduced from the former correlation functions: densities are discontinuities of the resolvents and resolvents are Stieljes transforms of densities. For example:
\beq \hat{w}_1(x)=\int \frac{\rho_1(x')}{x-x'}dx' \Longleftrightarrow \rho_1(x)=\frac{1}{2i\pi} \left(\hat{w}_1(x-i0)-\hat{w}_1(x+i0)\right)\eeq
Then we want to use a formal $\frac{1}{T}$ power-series development which unfortunately is not necessarily well-defined for all matrix models. Indeed, if one is interested in \textit{convergent matrix models}, then one must be sure that such a series expansion commutes with integrations. In general, this does not happen and solutions of the convergent matrix model differ from the solutions of the \textit{formal matrix model} (where by definition the development is assumed to exist and to commute with integrations). The explanation of this phenomenon is simple: when we use a series expansion, it automatically ignores the exponentially small factors (one can think, for example, to $\exp(-x^2)$ which has at $x=\infty$ the same asymptotic expansion as the zero function). To sum up, formal matrix models are easier to handle, because by definition the formal expansion exists and we can perform formal operations on it; but the price to pay is that we only get a part of the convergent solutions (we miss the exponentially decreasing terms). It could appear disappointing to consider just formal matrix models, since they do not carry the whole convergent solutions (and thus leads only to a significative but incomplete part of the convergent solutions), but fortunately differences between formal and convergent matrix models have been well studied, and in \cite{ConvForm1}, \cite{ConvForm2}, the authors show how to reconstruct with theta functions the convergent solutions from the formal ones. Moreover recently, in the article \cite{GBE}, the authors showed that once we assume the existence  of the asymptotic expansion for the resolvent and the two--points function (formulas (\ref{TopDev2}) below for $n=1,2$), then the same kind of expansion is guaranteed for all correlations functions and for the partition function as well. From now on, we will place ourselves in the case of formal matrix models, i.e. we assume that there automatically exists an expansion of type:
\beq \label{TopDev}\ln Z_N\mathop{=}^{\sim}\sum_{g=0}^\infty \left(\frac{N}{T}\right)^{2-2g}\hat{f}_g\eeq
and 
\beq \label{TopDev2} \hat{w}_n(x_1,\dots,x_n)=\sum_{g=0}^\infty \left(\frac{N}{T}\right)^{2-2g-n}\hat{w}_n^{(g)}(x_1,\dots,x_n)\eeq
The numbers $\hat{f}_g$ are called \textit{symplectic or spectral invariants} of the model (invariant relatively to symplectic transformations of the spectral curve). We refer the reader to the recent article \cite{BorEy} about the existence of such an expansion. The previous expansion can be understood as a large $N$ expansion and therefore in the limit $N\to \infty$ one expects that the leading value ($g=0$) corresponds to the ''real'' large $N$ limit of the model. In fact this intuition is correct and it has been proved that
\beq \hat{y}(x)=i\pi\rho_{eq}(x)=\frac{1}{2}V'(x)-\hat{w}_1^{(0)}(x).\eeq 
This formula establish a direct link between the equilibrium density and the leading order of the first correlation function. The function $\hat{y}(x)$ (which is up to a trivial rescaling the equilibrium density) is often named the \textit{spectral curve} of the problem. In our case, it satisfies:
\beq \hat{y}^2(x)=\text{Polynomial}(x)=\frac{1}{2T} h^2(x,T)(x-a_1(T))(x-b_1(T))(x-a_2(T))(x-b_2(T))\eeq
This identity defines the algebraic spectral curve $\hat{y}^2=P(x)$ where $P$ is a polynomial. We remind the reader that Eynard and Orantin showed in \cite{OE} that for any algebraic curve $P(x,y)=0$ we can associate some symplectic invariants $f_g$ and $w_n^{(g)}(x_1,\dots,x_n)$. Moreover, when the algebraic curve comes from a matrix model, these invariants are the same as the one we defined earlier in (\ref{Corrfunctions}) and (\ref{Corrfunctions2}). 

In our case, the function $\hat{y}(x,T)=\frac{1}{2T} h^2(x,T)(x-a_1(T))(x-b_1(T))(x-a_2(T))(x-b_2(T))$ depends on the temperature $T$ and so are the corresponding invariants $\hat{w}_n(x_1,\dots,x_n,T)$ and $f_g(T)$. When $T\to T_c$ it is known that $ \forall g>1, \, f_g\to \infty$ and that the correlation functions diverges. This is so because the expansion (\ref{TopDev}) reaches its radius of convergence in $T$. In order to recover finite quantities, one has to rescale properly the variables at $T\sim T_c$. In our case we will prove that the good rescaling is given by:
\beq x_i=b\epsilon+(T-T_c)^{\frac{1}{2m}}\xi_i\eeq
so that
\beq \hat{y}_{\text{rescaled}}(\xi):=\lim_{T\to T_c} \frac{\hat{y}( b\epsilon+(T-T_c)^{\frac{1}{2m}}\xi, T)}{T-T_c}\eeq
and
\beq \hat{w}_{\text{rescaled},n}^{(g)}(\xi_1,\dots,\xi_n):=\lim_{T\to T_c} \frac{\hat{w}_n^{(g)}( b\epsilon+(T-T_c)^{\frac{1}{2m}}\xi_1,\dots,b\epsilon+(T-T_c)^{\frac{1}{2m}}\xi_n,T)}{(T-T_c)^n}\eeq
and
\beq \hat{f}_{\text{rescaled},g}:=\lim_{T\to T_c} (T-T_c)^{-(2-2g)}\hat{f}_g\eeq
are finite quantities and that the new $\hat{w}_{\text{rescaled},n}^{(g)}(\xi_1,\dots,\xi_n)$ and $\hat{f}_{\text{rescaled},g}$ are the spectral invariants of the rescaled curve $\hat{y}_{\text{rescaled}}(\xi)$. In the general context of matrix model, such a rescaling is called a \textit{double scaling limit} since we have performed a double limit $N\to \infty$ and $T\to T_c$ so that $N(T-T_c)^{2m}$ remains finite:
\beq \ln Z_N=\sum_{g=0}^\infty \left(\frac{N}{T}\right)^{2-2g} \hat{f}_g\sim \sum_{g=0}^\infty \left(\frac{N}{T_c}\right)^{2-2g}(T-T_c)^{(2-2g)}\hat{f}_{\text{rescaled},g}\eeq
From a geometric point of view, this double scaling limit corresponds to a local zoom in the region of the degenerate point $b\epsilon$. The rate of the zoom depends on both the temperature $T$ and the size of the matrices $N$ so that $N(T-T_c)$ remains finite. It can be illustrated with the following picture:

\bigskip
\begin{center}
\includegraphics[height=6cm]{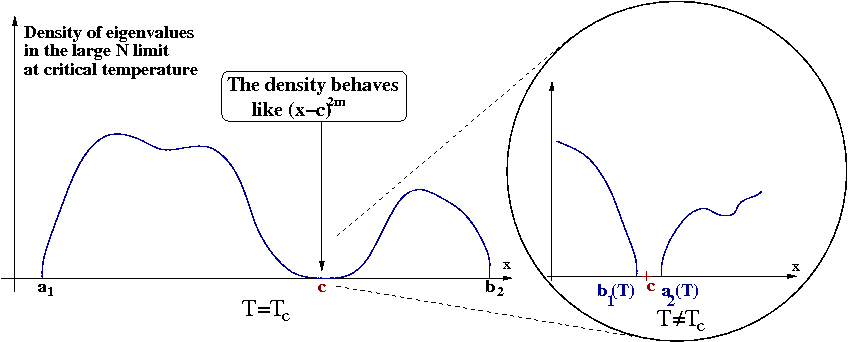}

	\underline{Figure 3}: Example of a critical eigenvalue density near the critical temperature
\end{center} 

In the context of matrix models, double scaling limits are often very important because they are expected to give universal (independent of the potential) rescaled spectral curve and correlation functions related to $(p,q)$ minimal models (and thus in our case the $(2m,1)$ minimal model). On the other hand, $(p,q)$ minimal models are studied through string reductions of some well known integrable systems. In the rest of the paper, we will prove that, in the case of the merging of two cuts, the rescaled spectral curve corresponds to the spectral curve of the $(2m,1)$ minimal model. Then, using the method introduced by Berg\`{e}re and Eynard in \cite{BergereEynard}, we prove that the rescaled correlation functions and the spectral invariants correspond to some ``correlation'' functions expressed with some  determinantal formulae \cite{determinantalformulae} for the $(2m,1)$ minimal model.

\subsection{The rescaled spectral curve in our $2m$ degenerate matrix model case} 

In order to get the rescaled spectral curve, we need to perform a few consecutive steps. First we can express explicitly the corresponding critical potential corresponding to $\rho_c(x)$ in \ref{SingularDensity} using \ref{Vconnexionh}.
The computation is straightforward and uses only the general Taylor expansion of:
\beq\sqrt{1+x}=1+\sum_{n=1}^\infty (-1)^{n+1} \frac{ (2n-2)!}{n!(n-1)!2^{2n-1}} x^n\eeq
It gives:
\beq\label{Potential} V'_c(x)=\sum_{j=0}^{2m+1} \left( \dbinom{2m}{j-1} (-b\epsilon)^{2m+1-j} +\sum_{n=1}^{E(\frac{2m+1-j}{2})} \dbinom{2m}{2n+j-1} \frac{(-1)^j(2n-2)!\epsilon^{2(m-n)+1-j}b^{2m+1-j}}{n!(n-1)!2^{2n-1}} \right) x^j 
\eeq
where $E(\frac{2m+1-j}{2})$ stands for the greatest integer lower or equal to $\frac{2m+1-j}{2}$. The critical temperature is given by:
\beq T_c=\frac{b^{2m+2}}{2}\sum_{n=1}^{m+1} \frac{\epsilon^{2m-2n+2}(2m)! }{n!(2m-2n+2)!(n-1)!2^{2n-1}}\eeq

Then, we need to use some reformulations of conditions \ref{ResidueConstraint} and \ref{hconnexionV}. Indeed, it is known for a long time (a proof can be found in appendix A of \cite{BergereEynard} but the results were derived much before) and has been used intensively in \cite{AAM} that the set of equations \ref{ResidueConstraint} and \ref{hconnexionV} leads to the following ordinary differential equations (sometimes called \textit{hodograph equations}):
\bea \label{Diffset}
     \frac{d }{dT}a_1(T)&=& \frac{4 (a_1(T)-x_0(T))}{h(a_1(T),T)(a_1(T)-b_1(T))(a_1(T)-a_2(T))(a_1(T)-b_2(T))}\cr
     \frac{d }{dT}b_1(T)&=& \frac{4 (b_1(T)-x_0(T))}{h(b_1(T),T)(b_1(T)-a_1(T))(b_1(T)-a_2(T))(b_1(T)-b_2(T))}\cr
     \frac{d }{dT}a_2(T)&=& \frac{4 (a_2(T)-x_0(T))}{h(a_2(T),T)(a_2(T)-b_2(T))(a_2(T)-a_1(T))(a_2(T)-b_1(T))}\cr
     \frac{d }{dT}b_2(T)&=& \frac{4 (b_2(T)-x_0(T))}{h(b_2(T),T)(b_2(T)-a_2(T))(b_2(T)-a_1(T))(b_2(T)-b_1(T))}\cr
\eea
where the point $x_0(T)$ ($b_1(T)\leq x_0(T)\leq a_2(T)$ ) is determined by:
\beq \int_{b_1(T)}^{a_2(T)} \frac{z-x_0(T)}{\sqrt{(b_1(T)-z)(z-a_1(T))(b_2(T)-z)(z-a_2(T))}}dz=0 \label{Definition x_0}\eeq 
This set of equations taken at $T=T_c$ for $a_1$ and $b_2$ gives:
\bea \frac{d a_1(T)}{dT}_{|T=T_c}&=&-\frac{2}{(1+\epsilon)^{2m}b^{2m+1}}\cr
\frac{d b_2(T)}{dT}_{|T=T_c}&=&\frac{2}{(1-\epsilon)^{2m}b^{2m+1}}\cr\eea
so that in a neighbourhood of $T_c$:
\beq \label{a1}\encadremath{a_1(T)\underset{T=T_c}{\sim}-b-\frac{2}{(1+\epsilon)^{2m}b^{2m+1}}(T-T_c)+o(T-T_c)}\eeq
\beq \encadremath{b_2(T)\underset{T=T_c}{\sim}b+\frac{2}{(1-\epsilon)^{2m}b^{2m+1}}(T-T_c)+o(T-T_c)} \eeq
As mentioned earlier, we expect that the functions $a_j(T)$ and $b_j(T)$ will be analytic functions of $\Delta=(T-T_c)^\nu$, where $\nu$ is an exponent that we will determine later. Therefore we introduce the following notations:
\bea  \label{Texpansion}
b_1(T)&=&b\epsilon+\alpha\Delta+ \sum_{n=1}^\infty b_{1,n}\Delta^n\cr
a_2(T)&=&b\epsilon+\gamma\Delta+ \sum_{n=1}^\infty a_{2,n}\Delta^n\cr
x_0(T)&=&b\epsilon+X_0\Delta+\sum_{n=1}^\infty x_{n}\Delta^n\cr
h(z,T)&=&T (z-b\epsilon)^{2m-1}+ P(z)\Delta+\sum_{n=1}^\infty h_{n}(z)\Delta^n\eea
where $P(z)$ and $h_n(z)$ are polynomials of degree at most $2m-2$. In equations \ref{Diffset} for $a_2(T)$ and $b_1(T)$ we see that the l.h.s. is of order $(T-T_c)^{\nu-1}$ whereas the r.h.s. is of order $(T-T_c)^{-(2m-1)\nu}$. Hence, to have compatible equations we must have, as announced in the previous subsection, that
\beq \label{Exponent} \encadremath{ \nu=\frac{1}{2m}}\eeq
The next step is purely technical and consists in proving that $\alpha=-\gamma$. Since it is only a technical point, we postpone this discussion in Appendix \ref{AppendixB}. With the help of this relation we can now determine the rescaled spectral curve.

First remember that for $T=T_c$ , we have \ref{SingularDensity}:
\beq (z-b\epsilon)^{2m}=h_c(z)=\underset{z\to \infty}{\text{Pol}} \left(\frac{V'_c(z)}{\sqrt{z^2-b^2}}\right) \label{Definitionhc}\eeq
For $T\neq T_c$, reminding that $V(z,T)=\frac{V(z)}{T}$ and that $\rho(z,T)$ is defined with a factor $\frac{1}{T}$ in \ref{Density} (which will cancel the one of $V(x,T)$) we have:
\beq h(z,T)=\underset{z\to \infty}{\text{Pol}} \left(\frac{V'(z)}{\sqrt{(z-a_1)(z-b_2)(z-a_2)(z-b_1)}}\right)\eeq
We now use the fact that up to order $\Delta^{2m-1}$, both $a_1$ and $b_2$ are respectively equal to $-b$ and $b$ (\ref{a1}). Therefore we get:
\beq h(z,T)=\underset{z\to \infty}{\text{Pol}} \left(\frac{V'(z)}{\sqrt{(z^2-b^2)(z-a_2)(z-b_1)}} +O\left(\Delta^{2m}\right)\right) \label{intermediate}\eeq
Then, from the definition of $h_c(x)$ we have that:
$$ (z-b\epsilon)^{2m}=\frac{1}{T_c}\underset{z\to \infty}{Pol} \left(\frac{V'(z)}{\sqrt{z^2-b^2}}\right)$$
so that:
\beq \frac{V'(z)}{\sqrt{z^2-b^2}}=T_c (z-b\epsilon)^{2m} + O\left(\frac{1}{z}\right)\eeq
Putting back this identity into \ref{intermediate} and noticing that $\frac{1}{\sqrt{(z^2-b^2)(z-a_2)(z-b_1)}}$ only gives negative powers of $z$ that will disappear when taking the polynomial part, we find that:
\bea h(z,T)&=&T_c \underset{z\to \infty}{\text{Pol}} \left(\frac{(z-b\epsilon)^{2m}}{\sqrt{(z-a_2)(z-b_1)}} +O\left(\Delta^{2m}\right)\right)\cr
&=& T_c \underset{z\to \infty}{\text{Pol}} \left(\frac{(z-b\epsilon)^{2m-1}}{\sqrt{1+\frac{2b\epsilon-a_2-b_1}{z-b\epsilon}+ \frac{(b\epsilon-a_2)(b\epsilon-b_1)}{(z-b\epsilon)^2}}} +O\left(\Delta^{2m}\right)\right)\cr
\eea
We can now insert the Taylor series of the square-root:
\beq (1+x)^{-\frac{1}{2}}=\sum_{n=0}^\infty \frac{(-1)^n(2n)!}{(n!)^22^{2n}}x^n\eeq
to get:
\bea  \frac{h(z,T)}{T_c}&=&\underset{z\to \infty}{\text{Pol}} \left((z-b\epsilon)^{2m-1}\left(1+\sum_{n=1}^\infty \frac{(-1)^n(2n)!}{(n!)^22^{2n}}\left(\frac{2b\epsilon-a_2-b_1}{z-b\epsilon}+ \frac{(b\epsilon-a_2)(b\epsilon-b_1)}{(z-b\epsilon)^2}\right)^n \right) \right)\cr
&&+O\left(\Delta^{2m}\right)\cr
&=&(z-b\epsilon)^{2m-1} +\cr
&&\underset{z\to \infty}{\text{Pol}} \Big(\sum_{n=1}^\infty\sum_{k=0}^n \frac{(-1)^n(2n)!}{(n!)^22^{2n}} \dbinom{n}{k}\left(2b\epsilon-a_2-b_1\right)^k \big((b\epsilon-a_2)(b\epsilon-b_1)\big)^{n-k} \cr
&&(z-b\epsilon)^{2m-1+k-2n} \Big) +O\left(\Delta^{2m}\right) \cr
\eea
Let's now introduce the following ensemble :\beq
\label{set}I_m=\{ (n,k)\in (\mathbb{N}^* \times \mathbb{N}) \, / \, 2n-k\leq 2m-1 \text{ and } k\leq n\}\eeq
Clearly $I_m$ is a finite set and we can rewrite the previous identity as:
 \bea  \frac{h(z,T)}{T_c}&=&(z-b\epsilon)^{2m-1} + \cr
&&\Big(\sum_{(n,k)\in I_m} \frac{(-1)^n(2n)!}{(n!)^22^{2n}} \dbinom{n}{k}\left(2b\epsilon-a_2-b_1\right)^k \big((b\epsilon-a_2)(b\epsilon-b_1)\big)^{n-k} \cr
&&(z-b\epsilon)^{2m-1+k-2n} \Big)+O\left(\Delta^{2m}\right) \cr
\eea
We can now introduce the series expansion in $\Delta$: 
$$2b\epsilon-a_2-b_1=-(\alpha+\gamma)\Delta+O\left(\Delta^2\right)$$ 
and
$$(b\epsilon-a_2)(b\epsilon-b_1)=\left(\alpha\Delta+\sum_{n=2}^\infty a_{2,n}\Delta^n\right)\left(\gamma\Delta+\sum_{n=2}^\infty b_{1,n}\Delta^n\right)=\Delta^2 \alpha\gamma+ O\left(\Delta^3\right)$$
Then we perform the rescaling \beq z=b\epsilon+\Delta\xi \label{rescaling}\eeq
We only need to take into account terms with degree strictly less than $\Delta^{2m}$ so that only a few terms remain:

\bea
\frac{h(\xi,\Delta)}{T_c}&=&\left( \xi^{2m-1}+ \sum_{(n,k)\in I_m} \frac{(-1)^n(2n)!}{(n!)^22^{2n}} \dbinom{n}{k}(-1)^k (\alpha+\gamma)^k (\alpha\gamma)^{n-k}\xi^{2m-1+k-2n} \right)\Delta^{2m-1}\cr
&&+O\left(\Delta^{2m}\right)
\eea
so that:
\beq h_{\text{rescaled}}(\xi)=T_c\left( \xi^{2m-1}+ \sum_{(n,k)\in I_m} \frac{(-1)^n(2n)!}{(n!)^22^{2n}} \dbinom{n}{k}(-1)^k (\alpha+\gamma)^k (\alpha\gamma)^{n-k}\xi^{2m-1+k-2n} \right)\eeq

Eventually we get the rescaled spectral curve by taking into account the trivial term $R^{1/2}(z,T)=\sqrt{(z-a_1(T))(z-a_2(T))(z-b_1(T))(z-b_2(T))}$ with the rescaling \ref{rescaling}:
\bea R^{\frac{1}{2}}(b\epsilon+\xi\Delta,\Delta)&=&\sqrt{(b\epsilon+\xi\Delta-a_1(\Delta))(b\epsilon+\xi\Delta-a_2(\Delta))(b\epsilon+\xi\Delta-b_1(\Delta))(b\epsilon+\xi\Delta-b_2(\Delta))}\cr
&=&b\sqrt{\epsilon^2-1}\sqrt{(b\epsilon+\xi\Delta-a_2(\Delta))(b\epsilon+\xi\Delta-b_1(\Delta))}+O\left(\Delta^{2m}\right)\cr
&=&ib\Delta\sqrt{1-\epsilon^2}\sqrt{(\xi-\alpha)(\xi-\gamma)}+O\left(\Delta^2\right)\cr
\eea
so that:
\bea \rho(b\epsilon+\xi\Delta,\Delta)&=& \frac{b\sqrt{1-\epsilon^2}}{2\pi}\sqrt{(\xi-\alpha)(\xi-\gamma)}\cr
&&\left( \xi^{2m-1}+ \sum_{(n,k)\in I_m} \frac{(-1)^n(2n)!}{(n!)^22^{2n}} \dbinom{n}{k}(-1)^k (\alpha+\gamma)^k (\alpha\gamma)^{n-k}\xi^{2m-1+k-2n} \right)\Delta^{2m} \cr
&&+O\left(\Delta^{2m+1}\right)\eea
giving that:
\bea \hat{y}_{\text{rescaled}}(\xi)&=&b\pi\sqrt{1-\epsilon^2}\sqrt{(\xi-\alpha)(\gamma-\xi)}\cr
&&\left( \xi^{2m-1}+ \sum_{(n,k)\in I_m} \frac{(-1)^n(2n)!}{(n!)^22^{2n}} \dbinom{n}{k}(-1)^k (\alpha+\gamma)^k (\alpha\gamma)^{n-k}\xi^{2m-1+k-2n} \right)\cr\eea
In the appendix \ref{AppendixB}, we prove that $\alpha=-\gamma$ so that it eventually leads to:
\beq \label{RescaledCurve}\encadremath{\alpha=-\gamma \,\, ,\,\, :\,  \hat{y}_{\text{rescaled}}(\xi)=b\pi\sqrt{1-\epsilon^2}\sqrt{(\gamma^2-\xi^2)}\left( \xi^{2m-1}+ \sum_{n=1}^{m-1} \frac{(2n)!}{(n!)^22^{2n}} \gamma^{2n}\xi^{2m-1-2n} \right)}\eeq

We can even compute the precise value of $\gamma$. Indeed, using \ref{set} to compute the leading term of the $\Delta$-expansion of $h(a_2,T)$ and putting it back into \ref{Diffset} (and using the fact that with the definition of $x_0$ \ref{Definition x_0} we have $X_0=0$ when $\alpha+\gamma=0$) we have:

\beq \encadremath{ \label{gamma}\alpha=-\gamma\,\,\, \text{with} \,\,\, \gamma^{2m}=\alpha^{2m}= -\frac{4m}{b^2(1-\epsilon^2)\left(\underset{n=0}{\overset{m-1}{\sum}} \frac{(2n)!}{(n!)^22^{2n}}\right) }=-\frac{(m!)^22^{2m+1}}{b^2(1-\epsilon^2)(2m)!} }\eeq 
In this case, introducing the new variable $s$ by $\xi=\gamma s$ or equivalently 
\beq z=b\epsilon+\gamma\Delta s\eeq
we get:
\beq \label{RescaledCurve2}\encadremath{\alpha=-\gamma \,\, ,\,\, :\,  \hat{y}_{\text{rescaled}}(s)=b\pi\gamma^{2m}\sqrt{1-\epsilon^2}\sqrt{(1-s^2)}\left( s^{2m-1}+ \sum_{n=1}^{m-1} \frac{(2n)!}{(n!)^22^{2n}} s^{2m-1-2n} \right)}\eeq

Eventually \ref{RescaledCurve2} shows as expected that when performing a double scaling limit $z=b\epsilon+\gamma\Delta s$ (with $\gamma$ a complex number given by \ref{gamma} whose argument gives oscillations in the ($\Re(z)$,$\Im(z)$) plane), we recover a universal curve. In the next section, we will see that this rescaled spectral curve  \ref{RescaledCurve} is exactly, (up to the trivial normalization factor $b\sqrt{1-\epsilon^2}$) the spectral curve arising in the Lax pair representation of the Painlev\'{e} II hierarchy with $t_m=1$, all other $t_j$'s (See next section for a definition) taken to zero and the identification $u_0(t)=\gamma$ (coherently with \ref{u0Equation}). Before proceeding in the study of the Lax pair representation, we remind the reader that from general results of Eynard and Orantin \cite{OE}, the rescaled invariants and correlation functions $\hat{w}_{\text{rescaled},n}^{(g)}$ and $\hat{f}_{\text{rescaled},g}$ are automatically the symplectic invariants and correlation functions of the new rescaled spectral curve $\hat{y}_{\text{rescaled}}(\xi)$ and thus do automatically satisfied the famous loop equations \cite{OE}.

\section{Correlation functions and invariants arising in the Lax pair representation of the $(2m,1)$ minimal model}

In the previous section, we have found the rescaled spectral curve for a double scaling limit of a $2m$ degenerate merging of two cuts in matrix models. As conjectured in \cite{BleherEynard}, we expect that this universal double scaling limit is connected to the Painlev\'{e} II hierarchy. In order to prove this result, we will follow the approach \cite{BergereEynard} developed and successfully applied for the $(2m+1,2)$ models. It consists in finding a natural spectral curve $y_{\text{Lax}}(x)$ from a Lax pair representation of the hierarchy and check that it is equal to our rescaled curve defined in the previous section. Then from another work of Berg\`{e}re and Eynard,\cite{determinantalformulae} we can define from the Lax pair representation some new correlation functions $W_n^{(g)}(x_1,\dots,x_n)$ and invariants $F_g$ by some determinantal formulae and a suitable kernel. In particular, they proved that these new functions do satisfy the same loop equations as our correlations functions. Eventually, with the study of the pole structure and $W_2^{(0)}$ we will end by proving that our new correlation functions $W_n^{(g)}(x_1,\dots,x_n)$ and invariants $F_g$ are identical to the rescaled ones defined in the previous section.

\subsection{A Lax pair representation for the $(2m,1)$ minimal model}

In their paper \cite{BleherEynard}, the authors claimed that a good Lax pair representation for the $(2m,1)$ minimal model should be given by a set of two $2\times 2$ matrices $\mathcal{R}(x,t)$ and $\mathcal{D}(x,t)$ satisfying the following Lax pair representation:
\bea \label{LaxPair}
\frac{1}{N}\frac{\partial}{\partial x} \Psi(x,t)&=& \mathcal{D}(x,t) \Psi(x,t)\cr
\frac{1}{N}\frac{\partial}{\partial t} \Psi(x,t)&=& \mathcal{R}(x,t) \Psi(x,t)
\eea
where $\Psi(x,t)$ is a two by two matrix whose entries will be written as:
\beq \Psi(x,t)=\begin{pmatrix}
                \psi(x,t) & \phi(x,t)\\
		\td{\psi}(x,t)& \td{\psi}(x,t)\\
               \end{pmatrix}
\eeq
and satisfies the normalization $\det\Psi(x,t)=1$.

The compatibility condition of the Lax pair is then:
\beq \label{CompatibilityCondition}
\left[\frac{1}{N} \frac{\partial}{\partial x} -\mathcal{D}(x,t), \mathcal{R}(x,t)-\frac{1}{N} \frac{\partial}{\partial t}\right]=0
\eeq
In order to specify completely the Lax pair, we need to impose some conditions about the shape of the matrices $\mathcal{R}(x,t)$ and $\mathcal{D}(x,t)$. In our case we will assume:
\beq 
\mathcal{R}(x,t)=\begin{pmatrix}
                  0 & x+u(t)\\
		  -x+u(t) &0\\
                 \end{pmatrix}
\eeq
and 
\beq 
\mathcal{D}(x,t)=\sum_{k=0}^m t_k \mathcal{D}_k(x,t)
\eeq
with
\beq
\mathcal{D}_k(x,t)=\begin{pmatrix}
                  -A_k(x,t) & xB_k(x,t)+C_k(x,t)\\
		  xB_k(x,t)-C_k(x,t) &A_k(x,t)\\
                 \end{pmatrix}
\eeq
and $A_k$, $B_k$, $C_k$ are polynomials of $x$ of degree respectively $2k-2$, $2k-2$, $2k$. Note that in the literature one can find several different Lax pair corresponding to the same problem. Indeed any conjugation (change of basis) give equivalent matrices that describe the same problem but in different coordinates (see section \ref{SectionLax}). In fact any equivalent Lax pair can be used since the quantities we will define later will be invariant from this choice.
In order to have more compact notation, we will use the following convention: a dot will indicate a derivative relatively to $t$ normalized by a coefficient $1/N$, namely:
\beq \label{Notation} \dot{f}(x,t)\mathop{=}^{\text{def}}\frac{1}{N} \frac{\partial f(x,t)}{\partial t}\eeq
Putting back this specific shape of matrices into the compatibility equation gives the following recursion:
\bea A_0&=&0, B_0=0, C_0=1\cr
C_{k+1}&=&x^2C_k+ \check{R}_k(u)\cr
B_{k+1}&=&x^2 B_k +\hat{R}_k(u)\cr
A_{k+1}&=&x^2 A_k+\frac{1}{2} \dot{\hat{R}}(u)
\eea 
where $\check{R}_k$ and $\hat{R}_k$ are the modified Gelfand-Dikii polynomials given by the following recursion:
\bea \label{recursionGD}
\hat{R}_0(u)&=&u \,\, \check{R}_0(u)=\frac{u^2}{2}\cr
\hat{R}_{k+1}(u)&=&u\check{R}_k(u)-\frac{1}{4}\frac{d^2}{d t^2} \hat{R}_k(u)\cr
\frac{d}{d t}\check{R}_k(u)&=&u \frac{d}{d t}\hat{R}_k(u)
\eea

It is then easy to see that the matrices $\mathcal{R}(x,t)$ and $\mathcal{D}(x,t)$ satisfy \ref{LaxPair} if and only if $u(t)$ satisfies the string equation (see details in \cite{BleherEynard}.)
\beq \label{StringEquation} \encadremath{
\sum_{k=0}^m t_k \hat{R}_k(u(t))=-tu(t)}\eeq
which gives an explicit differential equation of order $m$ satisfied by $u(t)$ (since the polynomials $\hat{R}_k$ can be explicitly computed from the recursion \ref{recursionGD}). In particular the case $m=1$ gives Painlev\'e II equation:
\beq \frac{d^2 u}{d t^2}(t)=2u^3(t)+4(t+t_0)u(t)\eeq
where $t_0$ is a free parameter that can be set to $0$ by a time-translation $\td{t}=t+t_0$.

\medskip

\underline{Remark}: Seculiar equations

As it is always the case for a linear differential equation, we can get a seculiar equation on $\psi(x,t)$ by combining the two components of the differential equation in $t$ given by \ref{LaxPair}. In our case, we find that both $\psi(x,t)$ and $\phi(x,t)$ are solution of the seculiar equation:
\beq \label{Seculiar Equation}
\ddot{\psi}(x,t) -\frac{\dot{u}(t) \dot{\psi}(x,t)}{x+u(t)}=\left(u^2(t)-x^2\right)\psi(x,t)\eeq 
which by a simple standard change of variable can be transformed into a Schrodinger-like equation.

\subsection{Large $N$ development}

From the fact that a dot derivative contributes with a factor $\frac{1}{N}$, it is easy to see from the string equation \ref{StringEquation} that $u(t)$ admits a series development at large $N$:
\beq u(t)=\sum_{j=0}^\infty \frac{u_j(t)}{N^{2j}}= u_0(t)+\frac{u_1(t)}{N^2}+\dots \label{UExpansion}\eeq

\underline{Note}: The fact that $u(t)$ admits such a development in $\frac{1}{N^2}$ and not $\frac{1}{N}$ comes from the fact that the modified Gelfand-Dikii polynomials $\hat{R}_k$'s are a sum of terms involving only even numbers of dots-derivatives (i.e. even power of $\frac{1}{N}$).

Putting back this expansion into the string equation \ref{StringEquation} and looking at the power of $N^0$ of the series gives us that $u_0(t)$ must satisfy the following algebraic relation:
\beq \encadremath{
\label{u0Equation}
-t=\sum_{j=1}^m t_j \frac{(2j)!}{2^{2j} (j!)^2} u_0(t)^{2j}}
\eeq

From that result, it is then easy to see that the matrices $\mathcal{R}(x,t)$ and $\mathcal{D}(x,t)$ also admit a large $N$ expansion:
\beq \mathcal{R}(x,t)=\begin{pmatrix}
                  0 & x+u_0(t)\\
		  -x+u_0(t) &0\\
                 \end{pmatrix} 
+\frac{1}{N^2}\begin{pmatrix}
                  0 & u_1(t)\\
		  u_1(t) &0\\
                 \end{pmatrix}
+\dots= \sum_{j=0}^\infty \frac{\mathcal{R}_j(x,t)}{N^{2j}}\eeq
and
\beq \mathcal{D}(x,t)= \sum_{j=0}^\infty \frac{\mathcal{D}_j(x,t)}{N^{j}}\eeq
where the first matrix can be explicitly computed:

\beq \label{MatriceD0} 
\mathcal{D}_0(x,t)=\begin{pmatrix} 0& t+B_0+C_0\\
                    -t +B_0-C_0 &0\\
                   \end{pmatrix}
\eeq
with
\bea \label{totoro}
B_0&=&\sum_{j=1}^m t_j\sum_{k=0}^{j-1} x^{2(j-k)-1}\frac{(2k)!}{2^{2k} (k!)^2}u_0(t)^{2k+1}\cr
C_0 &=&\sum_{j=1}^m t_j\left(x^{2j}+\sum_{k=1}^{j} x^{2(j-k)}\frac{(2k)!}{2^{2k} (k!)^2}u_0(t)^{2k}\right)
\eea

It should also be possible to find equations defining recursively the next matrices $\mathcal{R}_j(x,t)$ and $\mathcal{D}_j(x,t)$ by looking at the next orders in the series expansion. But since we will have no use of such results we do not mention them here.

\subsection{Spectral Curve attached to the Lax pair}

By definition, the spectral curve of a differential system like \ref{LaxPair} is given by $\det(\,yId-\mathcal{D}_0(x,t))=0$, that is to say by the large $N$ limit of the eigenvalues of the spectral problem (which we expect to give the large $N$ limit of our matrix model). Note in particular that this definition is independent of a change of basis (conjugation by a matrix). From all the previous results, we can compute this two by two determinant and get:
\beq
y^2=\left(B_0+C_0+t\right)\left(B_0-C_0 -t\right) =B_0^2- \left(C_0+t\right)^2
\eeq
Then, since we have $x B_0=u_0(t)\left(C_0+t\right)$ from \ref{u0Equation} and \ref{totoro} a straightforward computation gives the product as:
\beq \label{SpectralCurve}\encadremath{
y_{\text{Lax}}^2=P(x,t)=\left(u_0(t)^2-x^2\right)\left(\sum_{j=1}^m t_j\sum_{k=0}^{j-1}\frac{x^{2(j-k)-1}(2k)!}{2^{2k} (k!)^2}u_0(t)^{2k}\right)^2 }\eeq

In particular in the specific case where $\forall j<m:\,t_j=0, $ and $t_m=1$, we find that the spectral curve reduces to:
\beq \label{SpectralCurveRed}\encadremath{
 \forall j<m:\, t_j=0,\, t_m=1 \, \,\Rightarrow \, y_{\text{Lax}}(x)=\sqrt{u_0(t)^2-x^2}\sum_{k=0}^{m-1}\frac{x^{2(m-k)-1}(2k)!}{2^{2k} (k!)^2}u_0(t)^{2k} }\eeq

\textbf{As expected, with the identification $u_0(t)=\gamma$ we recover exactly the rescaled-spectral curve of our matrix model \ref{RescaledCurve}}.

\underline{Note:}
In \ref{SpectralCurve} we can see that the only simple zeros of $P(x,t)$ are at $x=\pm u_0(t)$. Moreover since the polynomial $P(x,t)$ is obviously even and that there is no constant term in $x$ in the sum, we get that $P(x,t)$ has a double zero at $x=0$ and has double roots at some points $\pm \lambda_i, \, i=1,\dots,m-1$

\subsection{Asymptotics of the matrix $\Psi(x,t)$}

The next step in the method of \cite{BergereEynard} is to determine an asymptotic of the functions $\psi(x,t)$ and $\phi(x,t)$. From the Schrodinger-like equation \ref{Seculiar Equation}, we have a BKW expansion:
$$\psi(x,t)=g(x,t)e^{N h(x,t)}\left(1+ \frac{\psi_1(x,t)}{N}+\frac{\psi_2(x,t)}{N^2}+\dots\right)$$
Putting back into the seculiar equation gives the following result:
\bea \label{Asymptotics}
\psi(x,t)&=&\frac{1}{\sqrt{2}} \left(\frac{u_0(t)+x}{u_0(t)-x}\right)^{\frac{1}{4}}e^{N\int^t\sqrt{u_0^2(t')-x^2}dt'}\left(1+\frac{\psi_1(x,t)}{N}+\dots\right)\cr
\phi(x,t)&=&-\frac{1}{\sqrt{2}} \left(\frac{u_0(t)+x}{u_0(t)-x}\right)^{\frac{1}{4}}e^{-N\int^t\sqrt{u_0^2(t')-x^2}dt'}\left(1+\frac{\phi_1(x,t)}{N}+\dots\right)\cr
\td{\psi}(x,t)&=&\frac{1}{\sqrt{2}} \left(\frac{u_0(t)-x}{u_0(t)+x}\right)^{\frac{1}{4}}e^{N\int^t\sqrt{u_0^2(t')-x^2}dt'}\left(1+\frac{\td{\psi}_1(x,t)}{N}+\dots\right)\cr
\td{\phi}(x,t)&=&\frac{1}{\sqrt{2}} \left(\frac{u_0(t)-x}{u_0(t)-x}\right)^{\frac{1}{4}}e^{-N\int^t\sqrt{u_0^2(t')-x^2}dt'}\left(1+\frac{\td{\phi}_1(x,t)}{N}+\dots\right)
\eea

One can easily check that at dominant order in $N$ the previous asymptotics gives $\det(\Psi(x,t))=1+O\left(\frac{1}{N}\right)$. The next step is to transform the integration over $t$ in the exponential as a integral over $x$ by using the property of the spectral curve. Indeed, the spectral curve defines a Riemann surface which can be parametrized locally by $x(z,t)$ and $y(z,t)$ where $z$ is a running point on the Riemann surface. Thus, the function $y$ can be seen as both a function of $(z,t)$ or $(x,t)$. In order to avoid confusion here, we will write differently the function when it is seen as a function of $(z,t)$ or as a function of $(x,t)$ (we put a tilda for the function in $(x,t)$ and keep $y$ for the function of $(z,t)$):
\beq \td{y}(x,t)= \sqrt{P(x,t)}=y(z(x,t),t)\eeq
Then, using standard chain rule derivation, one can compute:
\beq \frac{\partial y}{\partial z}\frac{\partial x}{\partial t}-\frac{\partial y}{\partial t}\frac{\partial x}{\partial z}=-\frac{\partial \td{y}}{\partial t} \frac{\partial x}{\partial z}\eeq
From the expression of the spectral curve \ref{SpectralCurve} (which gives explicitly $\td{y}(x,t)$)  one can compute $\frac{\partial \td{y}}{\partial t}$:
\bea \frac{\partial \td{y}}{\partial t}&=&\frac{x u_0(\partial_t{u}_0)}{\sqrt{u_0^2-x^2}}\left(\sum_{j=1}^m t_j\sum_{k=0}^{j-1}\frac{x^{2(j-1-k)}(2k)!}{2^{2k} (k!)^2}u_0(t)^{2k}\right) \cr
&&+x(\partial_t{u}_0) \sqrt{u_0^2-x^2}\left(\sum_{j=1}^m t_j\sum_{k=1}^{j-1}\frac{x^{2(j-1-k)}(2k)! 2k}{2^{2k} (k!)^2}u_0(t)^{2k-1}\right)\cr
&=&\frac{x(\partial_t{u}_0)}{\sqrt{u_0^2-x^2}}\sum_{j=1}^mt_j\frac{(2j)! (2j)}{2^{2j} (j!)^2}u_0(t)^{2j-1}\cr
&=&-\frac{x}{\sqrt{u_0^2-x^2}}
\eea
To get the last identity, we have used the string equation \ref{u0Equation} for $u_0(t)$. Therefore by introducing the parametrization:
\beq \label{Xpara} z^2=u_0(t)^2-x^2 \Leftrightarrow x^2=u_0(t)^2-z^2\eeq
one finds that:
\beq x'(z,t)=\frac{\partial x}{\partial z}=-\frac{\sqrt{u_0^2-x^2}}{x}\,\,\, , \,\,\, \frac{\partial x}{\partial t}=-\frac{u_0(\partial_t u_0)}{x} \eeq
so that eventually:
\beq \frac{\partial y}{\partial z}\frac{\partial x}{\partial t}-\frac{\partial y}{\partial t}\frac{\partial x}{\partial z}=-\frac{\partial \td{y}}{\partial t} \frac{\partial x}{\partial z}=-\frac{x}{\sqrt{u_0^2-x^2}}\frac{\sqrt{u_0^2-x^2}}{x}=-1\eeq
The last identity can be rewritten as:
\beq \label{PoissonBracket} \encadremath{\frac{\partial y}{\partial t}\frac{\partial x}{\partial z}-\frac{\partial y}{\partial z}\frac{\partial x}{\partial t}=1}\eeq
and interpreted as the remaining of a non-commutative structure of $[P,Q]=\frac{1}{N}$ in the limit $N \to \infty$ which in such situations often transform into a Poisson structure for $y(z,t) \leftrightarrow P$ and $x(z,t)\leftrightarrow Q$ by simply replacing the commutator with a Lie bracket: \beq\{y(z,t),x(z,t)\}=1 \label{CommutationStructure}\eeq

With the help of this structure, we can get a reformulation of the integral:
\beq \frac{\partial \td{y}}{\partial t}=\frac{1}{x'(z)}\eeq
hence:
\beq\frac{\partial \int^x \td{y}dx}{\partial t}=z\eeq
and
\beq \int^t \sqrt{u_0^2(t')-x^2}dt'=\int^t zdt  =\int^x \td{y}dx\eeq
Eventually we have the following large $N$ developments:
\bea \label{Asymptotics2}
\psi(x,t)&=&\frac{1}{\sqrt{2}} \left(\frac{u_0(t)+x}{u_0(t)-x}\right)^{\frac{1}{4}}e^{N\int^x\td{y}dx}\left(1+\frac{\psi_1(x,t)}{N}+\dots\right)\cr
\phi(x,t)&=&-\frac{1}{\sqrt{2}} \left(\frac{u_0(t)+x}{u_0(t)-x}\right)^{\frac{1}{4}}e^{-N\int^x\td{y}dx}\left(1+\frac{\phi_1(x,t)}{N}+\dots\right)\cr
\td{\psi}(x,t)&=&\frac{1}{\sqrt{2}} \left(\frac{u_0(t)-x}{u_0(t)+x}\right)^{\frac{1}{4}}e^{N\int^x\td{y}dx}\left(1+\frac{\td{\psi}_1(x,t)}{N}+\dots\right)\cr
\td{\phi}(x,t)&=&\frac{1}{\sqrt{2}} \left(\frac{u_0(t)-x}{u_0(t)-x}\right)^{\frac{1}{4}}e^{-N\int^x\td{y}dx}\left(1+\frac{\td{\phi}_1(x,t)}{N}+\dots\right)
\eea

\subsection{Kernels and correlation functions in the Lax pair formalism}

It was established in \cite{determinantalformulae} that one can define a kernel $K(x_1,x_2)$ and define from it (through determinantal formulae) some functions $W_n(x_1,\dots,x_n)$ that have nice properties. In particular the authors showed in \cite{determinantalformulae} that these functions do satisfy some loop equations and thus are likely to correspond to our matrix model correlation functions. Following \cite{determinantalformulae} we define the kernel by:
\beq \label{K}
K(x_1,x_2)=\frac{\psi(x_1)\td{\phi}(x_2)-\td{\psi}(x_1)\phi(x_2)}{x_1-x_2}\eeq
Then we define the (connected) correlation functions by:
\beq W_1(x)=\psi'(x)\td{\phi}(x)-\td{\psi}'(x)\phi(x)\eeq
\beq \label{defcorr} W_n(x_1,\dots,x_n)=-\frac{\delta_{n,2}}{(x_1-x_2)^2}- (-1)^n\sum_{\sigma=cycles} \prod_{i=1}^n K(x_{\sigma(i)},x_{\sigma(i+1)})\eeq
and eventually we define non-connected functions $W_{n,n-c}$ by determinantal formulae:
\beq \label{defnoncorr}W_{n,n-c}(x_1,\dots,x_n)=\mathop{det}^{'}(K(x_i,x_j))\eeq
where the notation $\mathop{det}^{'}$ means that the determinant is computed in the usual way as a sum over permutations $\sigma$ of products $(-1)^\sigma\prod_{i=1}^n K(x_i,x_{\sigma_i})$, except for terms when $i=\sigma(i)$ and  when $i=\sigma(j) \,, j=\sigma(i)$. In such cases, one must replace $K(x_i,x_i)$ by $W_1(x_i)$ and $K(x_i,x_j)K(x_j,x_i)$ by $-W_2(x_i,x_j)$. For additional details, we invite the reader to look at (\cite{determinantalformulae}) 

\medskip

As in our problem we will need the large $N$ developments of these functions, we introduce the notations:
\bea \label{NexpansionKernelsCorrFuncts}K(x_1,x_2)&=&K_0(x_1,x_2)e^{N\int_{x_2}^{x_1}\td{y}dx}\left(1+\sum_{g=1}^\infty N^{-g}K^{(g)}(x_1,x_2)\right)\cr
W_n(x_1,\dots,x_n)&=&\sum_{g=0}^\infty N^{2-2g-n}W_n^{(g)}(x_1,\dots,x_n)\cr
W_{n,n-c}(x_1,\dots,x_n)&=&\sum_{g=0}^\infty N^{n-2g}W_{n,n-c}^{(g)}(x_1,\dots,x_n)\eea
Then, we can insert all our previous results concerning the leading terms of the series expansion \ref{Asymptotics2},\ref{defcorr} and  \ref{NexpansionKernelsCorrFuncts}. It gives:
\beq K_0(x_1,x_2)=\frac{1}{2(x_1-x_2)}\left(\left(\frac{u_0+x_1}{u_0-x_1}\right)^{\frac{1}{4}}\left(\frac{u_0-x_2}{u_0+x_2}\right)^{\frac{1}{4}}+\left(\frac{u_0-x_1}{u_0+x_1}\right)^{\frac{1}{4}}\left(\frac{u_0+x_2}{u_0-x_2}\right)^{\frac{1}{4}}\right)
\eeq
\beq W_1^{(0)}(x)=\td{y}(x)\eeq
and
\beq W_2^{(0)}(x_1,x_2)=\frac{1}{4(x_1-x_2)^2}\left(-2+ \sqrt{ \frac{(u_0+x_1)(u_0-x_2)}{(u_0-x_1)(u_0+x_2)} }+\sqrt{ \frac{(u_0-x_1)(u_0+x_2)}{(u_0+x_1)(u_0-x_2)} } \right)\eeq 

In order to get rid of the square-roots in the expressions above, it is better to introduce a proper parametrization of our spectral curve \ref{SpectralCurve}. Let us define:
\beq \label{Para2} \encadremath{x=\frac{u_0}{2}\left( z+\frac{1}{z}\right)=\frac{u_0(z^2+1)}{2z}  \Leftrightarrow z=\frac{1+\sqrt{x^2-u_0^2}}{u_0}}\eeq
In particular, under such a change of variables we obtain several useful identities:
\bea \sqrt{\frac{u_0-x}{u_0+x}}&=&i\frac{z-1}{z+1}\cr
 u_0-x&=&-\frac{u_0(z-1)^2}{2z}\cr
 u_0-x&=&\frac{u_0(z+1)^2}{2z}\cr
 \sqrt{u_0^2-x^2}&=&\frac{iu_0}{2z}(z+1)(z-1)\cr
\frac{d x(z)}{dz}&=& \frac{u_0(z^2-1)}{2z^2}
\eea

\medskip

Eventually we can rewrite $W_2^{(0)}$ in terms of the new variable $z$:
\beq W_2^{(0)}(z_1,z_2)=\frac{4z_1^2z_2^2}{u_0^2 (z_1^2-1)(z_2^2-1)(z_1z_2-1)^2}\label{W20}\eeq

Although these functions have some interesting features, they still depend on the choice of coordinates on the Riemann surface defined by the spectral curve. Therefore, we introduce similarly to \cite{BergereEynard} and \cite{OE} the corresponding differential forms:

\beq \mathcal{W}_n^{(g)}(z_1,\dots,z_n)=W_n^{(g)}(x(z_1),\dots,x(z_n))x'(z_1)\dots x'(z_n) +\delta_{n,2}\delta_{g,0} \frac{x'(z_1)x'(z_2)}{(x(z_1)-x(z_2))^2} \eeq
These differentials are symmetric rational functions of all their variables. Moreover as proved in the crucial theorem \ref{Pole Structure} these functions only have poles at $z_i=\pm 1$ (except again $\mathcal{W}_2^{(0)}(z_1,z_2)$ which may have a pole at $x(z_1)=x(z_2)$). Eventually, a direct computation from \ref{W20} gives:
\beq \encadremath{\mathcal{W}_2^{(0)}(z_1,z_2)=\frac{1}{(z_2-z_1)^2} }\eeq

\subsection{Loop equations, determinantal formulae, pole structure and unicity}

The previous determinantal definitions may seem rather arbitrary, but as we mention before they have the interesting property (proved in \cite{determinantalformulae}) to satisfy the following loop equations.
\begin{theorem} Loop equations satisfied by the determinantal functions:
 \bea \label{loopnc} P_n(x;x_1,\dots,x_n)&=& W_{n+2,n-c}(x,x,x_1,\dots,x_n)\cr
&&+\sum_{j=1}^n \frac{\partial}{\partial x_j} \frac{W_n(x,x_1,\dots,x_{j-1},x_{j+1},\dots,x_n)-W_n(x_1,\dots,x_n)}{x-x_j}\cr\eea
is a polynomial of the variable $x$. The previous theorem is equivalently reformulated for the standard connected functions after projection on $N^{2-2g}$ by the sets of equations (valid for every $g\geq 0$):
\bea \label{loop} P^{(g)}_n(x;x_1,\dots,x_n)&=& \sum_{h=0}^g \sum_{I \subset J} W_{1+|I|}^{(h)}(x,I)W_{1+n-|I|}^{(g-h)}(x,J/I) \cr
&&+\sum_{j=1}^n \frac{\partial}{\partial x_j} \frac{W^{(g)}_n(x,J/\{x_j \} )-W^{(g)}_n(x_1,\dots,x_n)}{x-x_j}\cr\eea
is a polynomial of the variable $x$.
\end{theorem}

We emphasize again that loop equations are an essential step because it is well known in the matrix model world \cite{MehtaBook} that the correlation functions introduced in our first section do satisfy these loop equations. Unfortunately, loop equations generally admit several solutions encoded essentially in the unknown coefficients of the polynomial $P_n$. Therefore we need some additional results to get unicity. The first one deals with the pole structure:

\begin{theorem}\label{Pole Structure}  \underline{Pole Structure}:

The functions $z\to \psi_k(z,t)$ are rational functions with poles only at $z\in \{ \pm i,0,\infty\}$. The coefficients of these fractions depend on $u_0(t)$ and its derivatives. Hence the determinantal correlation functions $W_n^{(g)}$ are symmetric and rational functions in the variables $z_i$ with poles only at $z_i=\pm 1$.
\end{theorem}

\underline{Proof}: The last part of the theorem is obvious from the definitions as soon as the results regarding the $\psi_k(z,t)$'s are established. This proof is presented in Appendix \ref{AppPoleStructure} and is highly non-trivial. It uses the whole structure of integrability (i.e. the two differential equations \ref{LaxPair}) to eliminate other possible poles (at the other zeros of $y_{\text{Lax}}(x)$).

With the knowledge of the pole structure of the $W_n^{(g)}$, the fact that they satisfy the loop equations and the knowledge of $W_2^{(0)}$ we have a unicity theorem. In fact under these conditions we can identify our differentials $\mathcal{W}_{n}^{(g)}$'s with the ones defined by the standard recursion relation introduced by Eynard and Orantin in \cite{OE}:

\begin{theorem}
The differentials $\mathcal{W}_{n}^{(g)}$ satisfy the following recursion:
\bea \label{recursion}
  \mathcal{W}_{n+1}^{(g)}(z_1,\dots,z_n,z_{n+1})&=&\Res_{z \to \pm 1} \frac{dz}{2u_0 y(z)(1-\frac{z_{n+1}}{z})(\frac{1}{z}-z_{n+1})}\big[ \mathcal{W}_{n+2}^{(g-1)}(z,\overline{z},z_1,\dots,z_n)\cr
&&+\sum_{h=0}^g \mathop{\sum}_{I \in J}^{'} \mathcal{W}_{1+|I|}^{(h)}(z,I)\mathcal{W}_{n+1-|I|}^{(g-h)}(\overline{z},J/I)\big]\cr
\eea
where $J$ is a short-writing for $J=(z_1,\dots,z_n)$ and $\overset{g}{\underset{{h=0}}{\sum}} \overset{'}{\underset{I \in J}{\sum}}$ means that we exclude the terms $(h,I)=(0,\emptyset)$ and $(h,I)=(g,J)$ in the sum. The notation $\overline{z}$ stands for the conjugate point of $z$ near the poles where the residue is taken. In our case: $\overline{z}=\frac{1}{z}$
\end{theorem}

\underline{Note}: It is worth noticing that in Eynard and Orantin's notation we have in our case (we omit the dependance in the $t$ parameter):
\bea \om(z)&=&y(z)\frac{u_0 (z^2-1)}{z^2}\cr
 y(\overline{z})&=&y(1/z)=-y(z)\cr
dE_z(p)&=&\frac{1}{2}\int_z^{\frac{1}{z}} \frac{ds}{(s-p)^2}=\frac{1-z^2}{2(z-p)(pz-1)}\eea
so that:
\beq
\frac{dE_z(z_{n+1})}{\om(z)}=\frac{z^2}{2u_0 y(z)(z-z_{n+1})(1-z_{n+1}z)}=\frac{1}{2u_0 y(z)(1-\frac{z_{n+1}}{z})(\frac{1}{z}-z_{n+1})}\eeq
 
\underline{Proof of \ref{recursion}}: The unicity proof has been done in various article but for completeness we rederive it here with our notations. First of all Cauchy's theorem states that:
\beq \mathcal{W}_{n+1}^{(g)}(z_1,\dots,z_{n+1})= \Res_{z \to z_{n+1}} \frac{dz}{z-z_{n+1}}\mathcal{W}_{n+1}^{(g)}(z_1,\dots,z_{n+1})\eeq
We can move the integration contour to enclose all other poles, i.e. only $\pm 1$ in our case:
\bea \mathcal{W}_{n+1}^{(g)}(z_1,\dots,z_{n+1})&=& \Res_{z \to \pm 1} \frac{dz}{z_{n+1}-z}\mathcal{W}_{n+1}^{(g)}(z_1,\dots,z_{n+1})\cr
&=&\Res_{z \to \pm 1} \frac{x'(z)dz}{z_{n+1}-z}W_{n+1}^{(g)}(x(z_1),\dots,x(z_{n+1}))\eea
We observe that the loop equations \ref{loop} can be rewritten in accordance with \ref{loop} in the following form by isolating the coefficients $W_1^{(0)}$ in the sum:
\bea
-2W_1^{(0)}(x)W_{n+1}^{(g)}(x_1,\dots,x_n,x)&=&\sum_{h=0}^g \mathop{\sum}^{'}_{I \subset J} W_{1+|I|}^{(h)}(x,I)W_{1+n-|I|}^{(g-h)}(x,J/I)\cr
&&+\sum_{j=1}^n \frac{\partial}{\partial x_j} \frac{W^{(g)}_n(x,J/\{x_j \} )-W^{(g)}_n(x_j,J/\{x_j \})}{x-x_j}\cr
&&-P_n^{(g)}(x;x_1,\dots,x_n)\eea
Then putting this back into the residue computation, observing that the polynomial $P_n^{(g)}(x;x_1,\dots,x_n)$ does not contribute to the residue, and using the relation between $x$ and $z$ we are left with \ref{recursion}.

\section{Lax pairs for the (2m,1) minimal model and for the Painlev\'e II hierarchy \label{SectionLax}}

\subsection{The (2m,1) minimal model and the Flashka-Newell Lax pair}

As observed in \cite{BleherEynard} the string equation \ref{StringEquation} is nothing but the $m^{\text{th}}$ member of the so-called Painlev\'e II hierarchy. The Painlev\'e II (PII) hierarchy, a collection of ODEs of order $2m$, arises as a self-similar reduction of the mKdV hierarchy. In the papers \cite{CJM} and \cite{MazzoccoMo} this relationship has been used to construct a Lax pair for the PII hierarchy starting from the relevant Lax pair for the modified KdV hierarchy. We call this PII Lax pair the Flashka-Newell Lax pair since the first member of the hierarchy was find, for the first time, in \cite{FN}. In this subsection we prove that, up to a linear transformation of the wave function and a rescaling of the variables, the Flashka-Newell Lax Pair is equivalent to the $(2m,1)$ minimal model Lax pair. In order to simplify notation we forget, in this section, the rescaling given by $1/N$ over the variables $x$ and $t$. We begin with the case $t_1=0=t_2=\ldots=t_{m-1}$. 

\begin{proposition}
	Define $\tilde\Psi$ as a new wave function
	$$\tilde\Psi:=J\Psi$$ with
	$$J:=\begin{pmatrix}
			1 & i\\
			i & 1
		\end{pmatrix}$$
	and set $t_m\longmapsto (4^{m+1}/2)$ (all other parameters $t_j$ equal to $0$). Then $\tilde\Psi$ satisfies the Flashka-Newell Lax pair as written in \cite{CJM}.
\end{proposition}
\underline{Proof}
 Since $J$ is constant we observe that $\tilde\Psi$ solve the Lax system
 \bea\label{FNLAxPair}
	\frac{\partial}{\partial x}\tilde{\Psi}(x,t)&=& \tilde{\mathcal{D}}_m(x,t) \tilde{\Psi}(x,t)\cr
	\frac{\partial}{\partial t} \tilde{\Psi}(x,t)&=& \tilde{\mathcal{R}}(x,t) \tilde{\Psi}(x,t)
\eea
 with $ \tilde{\mathcal{D}}(x,t), \tilde{\mathcal{R}}(x,t)$ obtained through conjugation with $J$; i.e.
 $$
 	\tilde{\mathcal{R}}(x,t)=J\mathcal{R}(x,t)J^{-1}=\begin{pmatrix}
											-ix & u\\
											u & ix
										\end{pmatrix}	
 $$
 and
 $$
 	\tilde{\mathcal{D}}_m(x,t)=\frac{4^{m+1}}{2}J\mathcal{D}_m(x,t)J^{-1}=\frac{4^{m+1}}{2}\begin{pmatrix}
					-iC_m(x,t) & iA_m(x,t)+xB_m(x,t)\\
					-iA_m(x,t)+xB_m(x,t) & iC_m(x,t).
				\end{pmatrix}
 $$
These two matrices are exactly the ones appearing in (16a) and (16b) in \cite{CJM} (modulo the identification $u\longleftrightarrow w, x\longleftrightarrow\lambda, t\longleftrightarrow z$). 
For the matrix $\tilde{\mathcal R}$ this is self-evident. For $\tilde{\mathcal{D}}_m$ we just have to observe that it has the same shape as the matrix written in the right-hand side of (16b) (see eqs (14); in particular the polar part in (16b) is zero thanks to (14b)). On the other hand this conditions, plus compatibility condition, determines uniquely $\tilde{\mathcal{D}}_m$.

\medskip

Of course the result above is extended to the case in which all $t_j$ enter in $\cal D$ just taking linear combinations of the matrices studied in the previous proposition. This has been done, for the Flashka-Newell pair, in \cite{MazzoccoMo} (note, nevertheless, that there the spectral parameter is rotated; $\lambda\rightarrow -i\lambda$). Hence we have the following proposition.

\begin{proposition}
	Under a rescaling of all time variables $t_j\longrightarrow \frac{4^{j+1}}{2} t_j$ the (2m,1)-minimal model Lax pair is equivalent to the Flashka-Newell Lax pair for the PII hierarchy.
\end{proposition}

\section{Conclusion and outlooks}

In section $1$, we have established that the double scaling limit of a matrix model with a $2m$-degenerate point can define a universal rescaled spectral curve $\hat{y}_{\text{rescaled}}(x)$. In section $1$ we also reminded that the correlation functions and symplectic invariants $\hat{w}_n^{(g)}(x_1,\dots,x_n)$ and $\hat{f}_g$ can also be rescaled in a suitable way in order to give some new functions  $\hat{w}_{\text{rescaled,n}}^{(g)}(x_1,\dots,x_n)$  and new symplectic invariants  $\hat{f}_{\text{rescaled,g}}$ corresponding respectively to the correlation functions and symplectic invariants of the rescaled curve $\hat{y}_{\text{rescaled}}(x)$. Then, starting from a Lax pair of the Painlev\'{e} II hierarchy and using the same method as \cite{BergereEynard} we have constructed a spectral curve $y_{\text{Lax}}(x)$ which coincides with $\hat{y}_{\text{rescaled}}(x)$ for a natural choice of the flow parameters $t_j$'s. Finally, with the definition of a suitable kernel and determinantal formulae, we have defined in the same way as \cite{BergereEynard} some functions $W_n^{(g)}$ having interesting properties (loop equations). Studying in details the pole structure and computing $\mathcal{W}_{2}^{(0)}(z_1,z_2)$, we have eventually shown that the function $W_n^{(g)}$'s are in fact exactly the correlation functions of the curve $y_{\text{Lax}}(x)$. Since the two spectral curves are the same, we have proved the statement:
\begin{theorem}
The correlation functions (and spectral curve) of the double scaling limit of a $2m$-degenerate merging of two cuts are the same as the
functions $W_n^{(g)}$ (and spectral curve) defined by determinantal formulae of the integrable Painlev\'{e} II hierarchy's kernel.
\end{theorem}

This result reinforces the links between double scaling limit in matrix models and integrable $(p,q)$ minimal models. With this new result and the one of \cite{BergereEynard}, the two models are shown to be identical for $(p=2m ,q=1)$ and $(p=2m+1,q=2)$ ($m \in \mathbb{N}^*)$. However even if this identity is expected to hold for every $(p,q)$, some complete proofs as the one presented here are still missing. Indeed, if our reasoning may seem easy to generalize for arbitrary value of $p$ and $q$, the crucial theorem \ref{loop} establishing that the functions $W_n^{(g)}$ coming from determinantal formulae do satisfy the loop equations (proved in \cite{determinantalformulae}) is only valid for $q\leq2$ at the moment. Therefore a good approach to the generalization for arbitrary value of $(p,q)$ could be to first extend this theorem for every $(p,q)$ and then to use the method presented here to extend the result.

Another approach could be to use this approach to study other integrable systems whose Lax pairs are known. Indeed, it is possible to perform the same method as the one presented here for any Lax pair. In particular, for every Lax pair, it would be interesting to analyse the associated spectral curve and the corresponding determinantal correlation functions.

\setcounter{section}{0}

\appendix{Pole structure for $\psi_k(z,t)$}\label{AppPoleStructure}

In order to use the unicity theorem \ref{loop} showing that the $W_n^{(g)}$'s are the expected correlation functions, we need to precise the pole structure of the function $\psi_k(x,t)$'s and $\phi_k(x,t)$'s from which they are defined. In order to determine the functions $\psi_k(x,t)$'s, one can insert the series expansion \ref{Asymptotics2} into the seculiar equations. Since the case $\psi_k(x,t)$ and $\phi_k(x,t)$'s are similar (they satisfy the same seculiar equation), we will focus only on the $\psi_k(x,t)$'s. The main issue of this appendix is that putting the large $N$ asymptotics of $\psi(x,t)$ \ref{Asymptotics2} into the seculiar equation a priori gives unwanted poles at the zeros of $y(x)$ for $\psi_k(x,t)$ that we need to rule out. It is the purpose of this appendix to explain how this can be done.

\subsection{Study of the differential equation in $t$}

From the fact that $u(t)$ satisfies the string equation we remind the reader that we have \ref{u0Equation}:
\beq t=-\sum_{j=1}^m t_j \frac{(2j)!}{2^{2j} (j!)^2} u_0(t)^{2j}=P_0(u_0)\eeq
From this, it follows that $\frac{du_0}{dt}$ is:
\beq  \frac{du_0}{dt}=\frac{1}{P_0'(u_0)}\eeq
Performing more derivations relatively to $t$ can give the derivatives of $u_0(t)$ to any order as a fraction whose denominator is always a power of $P_0'(u_0)$. For example:
\bea \frac{d^2u_0}{dt^2}&=&-\frac{P''_0(u_0)}{(P_0'(u_0))^3}\cr
 \frac{d^3u_0}{dt^3}&=&-\frac{P_0'''(u_0)}{(P_0'(u_0))^4}+3\frac{(P_0''(u_0))^2}{(P_0'(u_0))^5}\eea
and so on.

As a consequence, any power of any derivative of $u_0$ remains a rational function of $u_0$ with poles only at the roots of $P_0'(x)$. For example, expressions like $\frac{d u_0}{dt}\frac{d^3 u_0}{dt^3}+ \left(\frac{d u_0}{dt}\right)^2\frac{d^2 u_0}{dt^2}$ will be rational functions of $u_0$ with poles only at the roots of $P_0'(x)$.

Now, putting back the development of $u(t)=u_0(t)+\frac{u_2(t)}{N^2}+\frac{u_3(t)}{N^3}+\dots$ into the full string equation \ref{StringEquation} gives that any subleading order $u_k$ can be expressed as a rational function of $u_0$ with poles only at the roots of $P'_0(x)$.

Eventually, inserting the shape of the function $\psi(x,t)$ into the seculiar equation and evaluating the order $N^{-k}$ gives the following equation $\forall k\geq 2$:
\bea \label{psik witht}
\partial_t \psi_{k-1}&=&\frac{\partial_{t^2}g}{2gh} \psi_{k-2} -\frac{\partial_{t}g}{gh}\partial_t \psi_{k-2} -\frac{\partial_{t^2}\psi_{k-2}}{2h} +\frac{1}{2} \left(\frac{\partial_t u}{u+x}\right)_{k} +\frac{1}{2} \sum_{i=0}^{k-2} \left(\frac{\partial_t u}{u+x}\right)_{k-i}\psi_i\cr
&&+\frac{\partial_t g}{2gh} \left(\frac{\partial_t u}{u+x}\right)_{k-1}+\frac{1}{2} \sum_{i=0}^{k-2}\left(\frac{\partial_t u}{u+x}\right)_{k-1-i}\left(\psi_i \frac{g_t}{gh} +\frac{\partial_t \psi_i}{h}\right)\cr
&&+\frac{1}{2} \sum_{i=0}^{k-2} \left(u^2\right)_{k-i} \frac{\psi_i}{h}
\eea
where we have written in short:
\bea
\psi_0(x,t)&=&1\cr
h(x,t)&=&\sqrt{u_0(t)^2-x^2}\cr
g(x,t)&=&=\left(\frac{u_0(t)+x}{u_0(t)-x}\right)^{1/4}\cr \label{definitions}
\eea
and the notation $\left(\frac{\partial_t u}{u+x}\right)_{k}$ stands for the term in $N^{-k}$ in the expansion of $\frac{\partial_t u}{u+x}$. Note in particular that these terms can be expressed as a fraction with poles at $u_0(t)+x=0$ and at $P'(u_0(t))=0$ (the last are independent of $x$).
For example the first one is:
\beq \label{psi1 witht} \partial_t \psi_1(x,t)=\frac{ (\partial_t u_0)^2 x^2 (u_0-x)^{\frac{3}{2}}}{4(u_0+x)^{\frac{3}{2}}} +\frac{(u_0+x)^{\frac{1}{2}}}{(u_0-x)^{\frac{1}{2}}} u_2(t)\eeq
where remember that $u_2(t)$ can be expressed as a rational function of $u_0(t)$ whose poles are known to be only when $u_0(t)$ is at a root of $P_0'$ (and thus are independent of $x$). From this expression, it is clear that $\psi_1(x,t)$ may only have $x$-dependent singularities at $x=\pm{u_0}$ and at $x=\infty$.

\subsection{Study of the differential equation in $x$}

The technic presented in the previous subsection can be carried out for the differential equation in $x$. Starting with the second equation of the Lax pair \ref{LaxPair}:
\beq 
\frac{1}{N}\frac{\partial} {\partial x} \Psi(x,t) =
\begin{pmatrix}
                  -A(x,t) & xB(x,t)+C(x,t)\\
		  xB(x,t)-C(x,t) &A(x,t)\\
                 \end{pmatrix}
\Psi(x,t)
\eeq
we can derive another seculiar equation for both $\psi(x,t)$ and $\phi(x,t)$:
\bea \label{seculiar}
0&=&\frac{1}{N^2} \frac{\partial^2}{\partial x^2}\psi(x,t) -\frac{1}{N^2} \left(\frac{ \partial_x (xB+C)}{xB+C}\right)\frac{\partial}{\partial x}\psi(x,t) \cr
&&+\frac{1}{N} \left(\partial_x A -A \frac{ \partial_x (xB+C)}{xB+C} \right) \psi(x,t)-y^2(x,t)\psi(x,t)\eea 
where we have used that:
\beq \det(\Psi)=1 \Leftrightarrow y^2(x,t)=A(x,t)^2+x^2B(x,t)^2-C(x,t)^2\eeq
Note in particular in the last identity that the r.h.s. should have a large $N$ development whereas the l.h.s. $y(x)$ given by \ref{SpectralCurve} does not. Therefore, the l.h.s. must have vanishing subleading orders in $\frac{1}{N^k} ,\forall \, k>0$.

Moreover, reformulating \ref{MatriceD0} gives:
\bea\label{identity2}
 A_0&=&0\cr
\left(xB+C\right)_0&=&y(x,t)\sqrt{\frac{u_0+x}{u_0-x}}
\eea
where the subscript $0$ stands for the first order in the large $N$ expansion. Indeed, it comes from the fact that:
\bea y^2(x,t)&=&\left(xB+C\right)_0\left(xB-C\right)_0=P(x,t)=P_1(x,t)P_2(x,t)\cr
 \left(xB+C\right)_0=P_1(x,t)&=&(u_0+x)\left(\sum_{j=1}^m t_j\sum_{k=0}^{j-1}\frac{x^{2(j-1-k)}(2k)!}{2^{2k} (k!)^2}u_0(t)^{2k}\right)\cr
 \left(xB-C\right)_0=P_2(x,t)&=&(u_0-x)\left(\sum_{j=1}^m t_j\sum_{k=0}^{j-1}\frac{x^{2(j-1-k)}(2k)!}{2^{2k} (k!)^2}u_0(t)^{2k}\right)\cr
\eea
and eventually:
\beq P_2(x,t)=P_1(x,t) \frac{u_0-x}{u_0+x} \eeq
With \ref{identity2} it is easy to see that:
\beq \label{identity3} \left(\frac{ \partial_x (xB+C)}{xB+C}\right)_0=\frac{\partial_x y}{y} +\frac{u_0}{u_0^2-x^2}\eeq
which will be crucial for the coherence of the computation. Indeed, putting the large $N$ expansion of $\psi(x,t)$:
$$ \psi(x,t)=g(x,t)e^{N h(x,t)}\left(1+ \frac{\psi_1(x,t)}{N}+\frac{\psi_2(x,t)}{N^2}+\dots\right)$$
into \ref{seculiar} and comparing the first orders in $\frac{1}{N}$ gives:
\bea 
0&=&g(x,t)y^2(x,t)-g(x,t)y^2(x,t)\cr
0&=& \partial_x(g(x,t)y(x,t)) +y(x,t)\partial_xg(x,t) +g(x,t)y(x,t) \left(\frac{ \partial_x (xB+C)}{xB+C}\right)_0
\eea
The second equation with the help of \ref{identity3} determines $g(x,t)$ coherently with \ref{definitions}, that is to say:
$$g(x,t)=\left(\frac{u_0(t)+x}{u_0(t)-x}\right)^{1/4}$$
Note now that $\forall \,k>0$, the function $\left(\frac{ \partial_x (xB+C)}{xB+C}\right)_k$ only has singularities at the singularities of $\frac{ 1}{xB_0+C_0}$ according to the standard rules of Taylor series for a fraction. 
The next order, $\frac{1}{N^2}$, gives us the function $\psi_1(x,t)$ (with the notation that a subscript $k$ defines the term in $N^{-k}$ in the expansion at large $N$):
\bea \label{psi1 withx}
\partial_x \psi_1(x,t)&=&-\frac{\partial_{x^2}g}{2gy}+ \frac{\partial_xg}{2gy}\left(\frac{ \partial_x (xB+C)}{xB+C}\right)_0+\frac{1}{2}\left(\frac{ \partial_x (xB+C)}{xB+C}\right)_1\cr
&&-\frac{1}{2}\left(\partial_xA -A\frac{ \partial_x (xB+C)}{xB+C}\right)_1\cr
\eea
From the definition of $g(x,t)$, it is easy to compute:
\bea \frac{\partial_x g}{g}&=&\frac{1}{2}\frac{u_0}{u_0^2-x^2}\cr
\frac{\partial_{x^2}g}{g}&=&\frac{u_0x}{u_0^2-x^2}+\frac{1}{4} \frac{u_0^2}{(u_0^2-x^2)^2}\eea
and thus to see that $\partial_x \psi_1(x,t)$ is a function of $x$ that may only have singularities at $x=\pm u_0$, at $x=\infty$ and at the others zeros of $y(x)=0$. (it is so because $\left(\frac{ \partial_x (xB+C)}{xB+C}\right)_1$  have the same singularities as $\frac{1}{xB_0+C_0}$ which by \ref{identity2} are only at $x=\pm u_0$, $x=\infty$ and at the zeros of $y(x)$).

It is then possible to extend this result for higher terms in the large $N$ expansion. The power $\frac{1}{N^k}$ gives:
\bea \label{psik withx}
\partial_x \psi_{k-1}&=&-\frac{\partial_{x^2}g}{2gy}\psi_{k-2}-\frac{\partial_xg}{gy}\partial_x \psi_{k-2} -\frac{1}{y}\partial_{x^2}\psi_{k-2}\cr
&&+\frac{1}{2}\sum_{i=0}^{k-2}\left(\frac{ \partial_x (xB+C)}{xB+C}\right)_{k-1-i} \psi_i +\frac{\partial_x g}{2gy}\sum_{i=0}^{k-2}\left(\frac{ \partial_x (xB+C)}{xB+C}\right)_{k-2-i}\psi_i\cr
&&+\frac{1}{2y}\sum_{i=0}^{k-2}\left(\frac{ \partial_x (xB+C)}{xB+C}\right)_{k-2-i}\partial_x\psi_i-\frac{1}{2y}\sum_{i=0}^{k-1-i}\left(\partial_x A- A\frac{ \partial_x (xB+C)}{xB+C}\right)_{k-1-i}\psi_i\cr
\eea
where we have defined $\psi_0=1$. The precise form of the relation is mostly irrelevant, but the main fact is that if all the $\psi_i(x,t) $ with $i<k$ are assumed to have singularities only at $x=\pm u_0$, $x=\infty$ and at the other zeros of $y(x)=0$, then the same is true for $\partial_x \psi_{k}$ by a simple recursion.

\subsection{Pole structure of $\psi_k(x,t)$}

With the help of \ref{psik witht} and \ref{psik withx} we are now able to prove that the only singularities of $x \mapsto \psi_k(x,t)$ are at $x=\pm u_0$ and at $x=\infty$. 

From \ref{psik witht} we have shown that $\partial_t \psi_k(x,t)$ can only have singularities at $x=\pm u_0(t)$, at $x=\infty$ and when $u_0(t)$ is at a root of $P_0'$. But from \ref{psik withx} we have shown that $\partial_x \psi_k(x,t)$ can only have singularities at $x=\pm u_0(t)$, at $x=\infty$ and at the other zeros of $y(x)=0$ given by $x=\lambda_i(t)$ solution of $\overset{m}{\underset{j=1}{\sum}} t_j\overset{j-1}{\underset{k=0}{\sum}}\frac{x^{2(j-k)-1}(2k)!}{2^{2k} (k!)^2}u_0(t)^{2k}=0$ in \ref{SpectralCurve}. But these poles are incompatible with the former result. Indeed if $\psi_k(x,t)$ had a pole at $x=\lambda_i(t)$, then $\partial_t \psi_k(x,t)$ would also have a pole at $x=\lambda_i(t)$, but we have shown that the only $x$-dependent singularities of $\partial_t \psi_k(x,t)$ are at $x\pm u_0(t)$ or $x=\infty$ giving rise to a contradiction. \textbf{Therefore:
 $x \mapsto \psi_k(x,t)$ has only singularities at $x=\pm u_0$ (square-root poles) and $x=\infty$ (poles) and in particular has no pole at the other zeros of $y(x)=0$}. This result is highly non trivial because we need to combine the two differential equations (i.e. the whole integrable structure) to get it. Hence, the structure of integrability seems to play an important underlying role in the pole structure and we can hope that such a result could extend to every integrable system. 

\subsection{Pole structure in the $z$ variable}

In order to have only poles (and not square root singularities), we want to shift the former result to the $z$ variable defined by:
\beq z^2=\frac{u_0-x}{u_0+x} \Leftrightarrow x=u_0 \frac{1-z^2}{1+z^2}\eeq
Note that we have the identities:
\bea \frac{\partial x}{\partial t}&=&(\partial_t u_0)\frac{1-z^2}{1+z^2}\cr
\frac{\partial x}{\partial z}&=&-\frac{4u_0z}{(1+z^2)^2}\cr
u_0+x&=&\frac{2u_0}{1+z^2}\cr
u_0-x&=&\frac{-2z^2}{1+z^2}\cr
g(z,t)&=&\frac{(-u_0)^{\frac{1}{4}}}{z^{\frac{1}{2}}}\cr
y(z,t)&=&\frac{4z^2u_0^2}{(1+z^2)^2} P_0(\left(\frac{1-z^2}{1+z^2}\right)u_0)\cr
\frac{\partial_x g}{g}(z,t)&=&\frac{(1+z^2)^2}{8u_0^2z^2}\cr
\frac{\partial_{x^2}g}{g}(z,t)&=&\frac{(1+z^2)^2}{4u_0^2z^2}+\frac{(1+z^2)^4}{64u_0^2z^4}\label{computations}\eea

Note also that every polynomial in $x$ will give a polynomial in $\frac{1-z^2}{1+z^2}$, that is to say a rational function in $z$ with poles at $z^2+1=0$.

The rules for derivation give that:
\beq \partial_t \td{\psi}_k(z,t)=\partial_t \psi_k(x,t)+\frac{\partial x}{\partial t} \frac{\partial \psi_k(x,t)}{\partial x}\eeq
\beq \partial_z \td{\psi}_k(z,t)=\frac{\partial x}{\partial z}\partial_x \psi_1(x,t) \eeq
where all these terms are already known from the previous sections. If one uses \ref{computations} and the remark that a polynomial in $x$ will give a rational function in $z$ with poles at $z^2+1=0$ (and remember that functions $A,B,C$ are polynomials in $x$), one can see that the singularities of $\psi_k(x,t)$ at $x=\pm{u_0}$ (square-root type) and at $x=\infty$ (poles), will transform into poles at $z=0$ ($\Leftrightarrow x=-u_0$), $z=\infty$ ($\Leftrightarrow x=u_0$) and $z=\pm i$ ($\Leftrightarrow x=\infty$).

Hence we have the final result:
\textbf{ \label{poles} $\forall k\geq 0$: the functions $z\to \psi_k(z,t)$ are rational functions with poles only at $z\in \{ \pm i,0,\infty\}$. The coefficients of these fractions depend on $u_0(t)$ and its derivatives.}  

\appendix{Discussion about $\alpha=-\gamma$}\label{AppendixB}

When computing the rescaled spectral curve in the matrix model double scaling limit, we need to find a relationship between $\alpha$ and $\gamma$ that are given by \ref{Texpansion}:
\bea
b_1(T)&=&b\epsilon+\alpha\Delta+ \sum_{n=1}^\infty b_{1,n}\Delta^n\cr
a_2(T)&=&b\epsilon+\gamma\Delta+ \sum_{n=1}^\infty a_{2,n}\Delta^n\cr
\eea
where we remind that $\Delta=(T-T_c)^{\frac{1}{2m}}$.
A first argument in favour of the fact that $\alpha=-\gamma$ is the case when $\epsilon=0$. Indeed, in such a case, the situation is fully symmetric around the singular point $0$. Therefore, one expects the two endpoints $b_1(T)$ and $a_2(T)$ to be symmetric around $x=0$ for every value of $T$ around $T_c$. In such a case the identity $\forall T \simeq T_c: \, a_2(T)=-b_1(T)$ gives $\alpha=-\gamma$. When $\epsilon\neq0$, we can carry out a similar reasoning at first orders in $\Delta$. Indeed, if we center the origin at $b\epsilon$, then as we observed it several times, the endpoints $a_1$ and $b_2$ can be considered to be respectively $-b$ and $b$ up to order $\Delta^6$. Therefore in the function $R^{\frac{1}{2}}(x)$ they only add a multiplicative trivial factor depending on $\epsilon$ ($\sqrt{1-\epsilon^2}$ to be precise) which will not change the symmetry around $b\epsilon$ of the endpoints $a_2$ and $b_1$ at first orders in $\Delta$.

Eventually, another more explicit approach is to put the developments \ref{Texpansion} into all the equations \ref{hconnexionV}, \ref{ResidueConstraint}, \ref{IntegralsConstraints} and \ref{Definition x_0} determining $h(z,T)$, $x_0(T)$ and the endpoints $a_1(T), b_1(T), a_2(T)$ and $b_2(T)$. Doing so leads to an algebraic equation of degree $2m$ connecting $\alpha$ and $\gamma$:
\beq Q(\alpha,\gamma)=0 \eeq
with $Q$ a symmetric, homogeneous polynomial of degree $2m$. Unfortunately the system does not admit a unique solution as soon as $m>1$. Indeed, although the solution $\alpha=-\gamma$ is always there, when $m>1$ there are also other possibilities such as $\alpha=\lambda \gamma, \lambda \in \mathbb{C}$ and $\gamma$ satisfying an equation of degree $2m$ with complex coefficients. Though it might appear surprising that the set of equations may have several distinct solutions (thus giving several eigenvalues density), one must remember that they are some additional constraints for the solution. Indeed, if one wants to have a density distribution, it means that all quantities involved must at least be real and positive. Therefore only the solution $\alpha=-\gamma$ is possible.

\underline{Note}: In fact $\alpha$ and $\gamma$ are not necessarily well defined. Indeed, they are only defined up to a multiplicative $(2m)^{\text{th}}$ root of unity since the equation defining them is homogeneous of degree $2m$. This is because the notion of $\Delta=(T-T_c)^{\frac{1}{2m}}$ is also ambiguous, whereas $\Delta^{2m}$, $\alpha^{2m}$ and $\gamma^{2m}$ are well-defined quantities. (which explain why the development in $a_1(T)$ and $b_2(T)$ is well defined). Indeed, if one changes: 
\beq \forall n \in \{ 1,\dots,2m-1 \}: \,\, \Delta \rightarrow \td{\Delta}=\Delta e^{\frac{2in\pi}{2m}} \, \, \, \text{,} \,\,\,  \alpha \rightarrow \td{\alpha}=\alpha e^{\frac{-2in\pi}{2m}}\, \, \, \text{and} \,\,\, \gamma \rightarrow \td{\gamma}=\gamma e^{-\frac{2in\pi}{2m}}\eeq
then \ref{Texpansion} remains unchanged. With the change $\xi \rightarrow \td{\xi}=\xi e^{-\frac{2in\pi}{2m}}$, the rescaled spectral curve remains unchanged.

\bibliographystyle{plain}
%\bibliography{/Users/mattiacafasso/Documents/BibDeskLibrary.bib}

\def\cprime{$'$}

\end{document}